\documentclass[aps,prc,superscriptaddress,twoside,twocolumn,nofootinbib,showpacs,floatfix]{revtex4-1}
\usepackage{amsmath,amssymb}
\usepackage{lipsum}
\usepackage{graphicx}
\usepackage{subfigure}
\usepackage[colorlinks]{hyperref}
\pdfpagewidth=\paperwidth
\pdfpageheight=\paperheight
\allowdisplaybreaks
\usepackage{newtxtext, newtxmath}
\usepackage{braket}

\begin{document}

\title{Resonating group method for baryon-baryon interactions with unequal oscillator frequencies and its application to the $N\Delta$ system in a chiral quark model}

\author{Ke-Rang Song}
\affiliation{School of Nuclear Science and Technology, University of Chinese Academy of Sciences, Beijing 101408, China}

\author{Fei Huang}
\email[Contact author: ]{huangfei@ucas.ac.cn}
\affiliation{School of Nuclear Science and Technology, University of Chinese Academy of Sciences, Beijing 101408, China}

\date{\today}

\begin{abstract}
The resonating group method (RGM) is widely used to investigate baryon-baryon interactions at the quark level, typically under the assumption that the two baryons involved share an identical harmonic-oscillator frequency. In reality, however, when a specific interaction Hamiltonian is given, different baryons should have unequal oscillator frequencies due to distinct interaction potentials induced by their different quantum numbers. In this work, we develop a new quark-level RGM formalism for baryon-baryon systems with unequal oscillator frequencies, with the aim of providing a unified and consistent framework for describing both single-baryon properties and baryon-baryon interactions. We present the formalism for solving bound-state and scattering problems, with particular emphasis on constructing the wave functions of two-baryon systems with unequal oscillator frequencies. The proposed formalism is then applied to the $N\Delta$ system within a chiral SU(3) quark model, where the quark-quark interaction includes, in addition to the one-gluon exchange (OGE) and a phenomenological confinement potential, the nonet scalar and pseudoscalar meson exchanges arising from the spontaneous breaking of chiral SU(3) symmetry. The distinctive features of the newly developed formalism are demonstrated by comparing the results from the new formulation with those from traditional calculations.
\end{abstract}

\maketitle

\section{Introduction}

As hadrons are composed of quarks and gluons, understanding baryon-baryon $(BB)$ interactions from the quark level remains a fundamental objective in strong-interaction physics. The resonating group method (RGM), which describes microscopic interactions between clusters, is one of the most well-established approaches for studying $BB$ interactions in terms of quark and gluon degrees of freedom. Over the past few decades, RGM has been highly successful in reproducing nucleon-nucleon $(NN)$ scattering phase shifts and hyperon-nucleon $(YN)$ scattering cross sections within constituent quark models \cite{Oka:1980,Harvey:1981,Faessler:1982,Shimizu:1984,Straub:1988,Brauer:1990,Valcarce:1994,Zhang:1994,Zhang:1997,Entem:2000,Fujiwara:2001,Mota:2002,Dai:2003,Ping:2009}. Progress has also been made in understanding kaon-nucleon ($KN$) and antikaon-nucleon (${\bar K}N$) interactions within a chiral SU(3) quark model \cite{Huang:2004,Huang:2004-2,Huang:2005,Huang:2007,Huang:2008}.

In previous quark-model investigations of $BB$ interactions based on RGM, the internal orbital wave functions of baryons were typically chosen as $(0s)^3$ harmonic oscillator wave functions with a common oscillator frequency adopted for all baryons. This assumption was adopted mainly to simplify the separation of the center-of-mass (CM) motion in two-baryon systems and to facilitate the evaluation of RGM kernel matrix elements. However, in a realistic model with a specific interaction Hamiltonian, different baryons should generally possess different oscillator frequencies due to their distinct quantum numbers. Assuming the same oscillator frequency for all baryons leads to an inconsistency: the chosen Gaussian wave functions for single baryons are no longer eigenfunctions of the given Hamiltonian, or in other words, the computed Hamiltonian matrix elements do not correspond to the actual energies of single baryons, which should instead be obtained by minimizing the expectation value of the Hamiltonian with respect to the oscillator frequency within the Gaussian basis. 

More importantly, when such single-baryon wave functions are employed in RGM studies of $BB$ interactions, additional reaction channels often need to be introduced to lower the energies of the two-baryon systems. These channels, however, are not entirely physical---they partly serve to compensate for the inadequacy of the single-baryon internal wave functions as eigenfunctions of the Hamiltonian. In such cases, caution must be exercised in interpreting each channel as a physical component of the two-baryon system. In Refs.~\cite{ohta:1982,liu:1982}, Ohta et al. and Liu demonstrated that once the stability condition of the three-quark nucleon $(N)$ is enforced---that is, when a specific oscillator frequency is chosen such that the single $N$ is a solution of the given Hamiltonian---the ``hidden color'' channel no longer plays a significant role in the $NN$ interaction. It is therefore natural to expect that, for other $BB$ systems, a proper understanding of the channel-coupling effects unavoidably requires a consistent treatment of both single baryons and $BB$ systems.

On the other hand, in quark models, a reliable extraction of the parameters of quark-quark $(qq)$ interactions---such as coupling constants and cutoff parameters---is essential for correctly understanding $BB$ interactions at the quark level and for making credible predictions for dibaryons, hadronic molecules, multiquark states, and other exotic systems. Typically, the parameters of $qq$ interactions are determined by fitting data on single-baryon energies, $NN$ scattering phase shifts, and $NY$ scattering cross sections. However, when a common oscillator frequency is used for all single baryons, the wave functions of most single baryons are not eigenfunctions of the Hamiltonian. This causes the extracted $qq$ interactions to be unphysical, thereby undermining a proper understanding of $BB$ interactions and casting doubt on the reliability of predictions for other hadronic systems based on such interactions.

For instance, in previous quark model studies, the confinement potential was believed not to contribute noticeably to the interactions between two color-singlet baryons. However, this conclusion holds only when a common oscillator frequency is pre-chosen for all single baryons (see Sec.~\ref{Sec:discuss} for a detailed discussion). Another example concerns the one-gluon exchange (OGE) potential, which is considered one of the most important short-range $qq$ interactions. In earlier quark model studies, the OGE coupling constant $g_u$ for $u(d)$ quark was determined by fitting the $N$-$\Delta$ mass difference. The underlying logic is that, with a common oscillator frequency, the contributions from the kinetic energy and confinement potential for $N$ are the same as those for $\Delta$, so their mass difference arises only from the OGE---provided that possible one-boson exchange (OBE) effects are explicitly included. However, the $g_u$ extracted in this way is not rigorously consistent with the Hamiltonian, as the $\Delta$ is not an eigenstate under the condition that the $N$ mass is minimized by adjusting the confinement coupling. This inconsistency could undermine the reliability of such a parameter when applied to other hadronic systems, motivating the need for a self-consistent framework as developed here.

In Ref.~\cite{Huang:2018}, we made the first attempt to describe the energies of single baryons, the binding energy of the deuteron, and the $NN$ scattering phase shifts in a consistent manner within a chiral SU(3) quark model. Unlike previous quark model calculations, where the oscillator frequencies of Gaussian wave functions were treated as predetermined parameters and taken to be the same for all single baryons, we determined the oscillator frequencies by variational method when computing the energies of single-baryon ground states. This ensures that all single baryons are minima of the given Hamiltonian. The model parameters in the Hamiltonian were then adjusted to simultaneously match the calculated energies of octet and decuplet baryons, the binding energy of the deuteron, and the predicted $NN$ scattering phase shifts with their experimental values. The results showed that the calculated masses of the octet and decuplet baryon ground states, the binding energy of the deuteron, and the $NN$ scattering phase shifts up to a total angular momentum $J=6$ and the mixing parameters for the relevant coupled partial waves are in satisfactory agreement with experiment. Most importantly, it clearly demonstrated that the resulting oscillator frequencies for the considered single baryons are quite different. In particular, the oscillator frequencies for octet baryons are much larger than those for decuplet baryons. Note that in Ref.~\cite{Huang:2018}, we reported the size parameter $b_u$ of the Gaussian wave function for each baryon, which is related to the oscillator frequency $\omega$ via $b_u^2 = 1/\left(m_u \omega\right)$, with $m_u$ being the constituent mass of the $u(d)$ quark.

Our previous work \cite{Huang:2018} serves as a good starting point to achieve a consistent and unified description of single-baryon properties and baryon-baryon dynamics. To generalize the study of $NN$ interaction in Ref.~\cite{Huang:2018} to systems composed of two different baryons, one needs the RGM formalism for two-baryon systems with unequal oscillator frequencies. In the literature, the RGM for two nuclei at the hadron level with unequal oscillator frequencies has been discussed, e.g., in Refs.~\cite{Horiuchi:1977,kamimura:1974}. However, to the best of our knowledge, the quark-level RGM for two baryons with unequal oscillator frequencies has never been presented or applied. 

In the present work, we develop a quark-level RGM formalism for two-baryon systems with unequal oscillator frequencies. We present the formalism for solving both bound-state and scattering problems, with particular emphasis on the construction of wave functions for two-baryon systems with unequal oscillator frequencies. The formalism is then applied to the $N\Delta$ system within a chiral SU(3) quark model, whose interaction consists of the nonet scalar and pseudoscalar meson exchanges arising from the spontaneous breaking of chiral SU(3) symmetry besides the OGE and phenomenological confinement potentials. By comparing the results from this framework with those from the traditional approach---where the oscillator frequency of $\Delta$ is set equal to that of $N$---we analyze the distinctive features of the proposed formalism. Our approach provides a unified description of single-baryon properties and $BB$ interactions, and enables a more physically grounded determination of model parameters, thereby establishing a reliable foundation for future studies of $BB$ interactions as well as exotic hadronic states such as dibaryons and multiquark systems.

The paper is organized as follows. In the next section, we present a new formulation of RGM with unequal oscillator frequencies. In Sec.~\ref{Sec:chiral-QM}, we review the main aspects of the chiral SU(3) quark model. The results of applying the new RGM formalism to the $N\Delta$ system are shown and discussed in Sec.~\ref{Sec:discuss}. Finally, a summary and conclusions are drawn in Sec.~\ref{Sec:summary}.

\section{RGM for $BB$ system with unequal oscillator frequencies}  \label{Sec:RGM}

The RGM for $BB$ systems with equal oscillator frequencies has been extensively used in previous quark model calculations. However, the RGM for $BB$ systems with unequal oscillator frequencies has never been discussed or applied in the existing literature. Here, following the discussion of the RGM for two nuclei at the hadron level with unequal oscillator frequencies in traditional nuclear physics \cite{Horiuchi:1977,kamimura:1974}, we develop a quark-level RGM for $BB$ systems with unequal oscillator frequencies.

The Jacobi coordinates for a $BB$ system are defined in the usual way:
\begin{align}
\boldsymbol{\xi}_1 &= \boldsymbol{r}_2 - \boldsymbol{r}_1,  \label{Jacobi-i} \\[6pt]
\boldsymbol{\xi}_2 &= \boldsymbol{r}_3 - \frac{m_1\boldsymbol{r}_1 + m_2\boldsymbol{r}_2}{m_1+m_2},  \\[6pt]
\boldsymbol{\xi}_4 &= \boldsymbol{r}_5 - \boldsymbol{r}_4,  \\[6pt]
\boldsymbol{\xi}_5 &= \boldsymbol{r}_6 - \frac{m_4\boldsymbol{r}_4 + m_5\boldsymbol{r}_5}{m_4+m_5}, \\[6pt]
\boldsymbol{R}_{\rm AB} &= \boldsymbol{R}_{\rm A} - \boldsymbol{R}_{\rm B}, \\[6pt]
\boldsymbol{R}_{\rm cm} &= \frac{M_{\rm A}\boldsymbol{R}_{\rm A} + M_{\rm B} \boldsymbol{R}_{\rm B}}{M_{\rm A}+M_{\rm B}},  \label{Jacobi-f}
\end{align}
with
\begin{align}
M_{\rm A} &= m_1 + m_2 + m_3,  \\[6pt]
M_{\rm B} &= m_4 + m_5 + m_6,  \\[6pt]
\boldsymbol{R}_{\rm A} &= \frac{m_1 \boldsymbol{r}_1 + m_2 \boldsymbol{r}_2 + m_3 \boldsymbol{r}_3}{M_{\rm A}},  \\[6pt]
\boldsymbol{R}_{\rm B} &= \frac{m_4 \boldsymbol{r}_4 + m_5 \boldsymbol{r}_5 + m_6 \boldsymbol{r}_6}{M_{\rm B}},
\end{align}
where $\boldsymbol{r}_i$ and $m_i$ $(i=1$--$6)$ are the coordinate and mass of the $i$-th constituent quark, respectively. The Jacobi coordinates $\boldsymbol{\xi}_1$ and $\boldsymbol{\xi}_2$ are the internal coordinates for cluster A, while $\boldsymbol{\xi}_4$ and $\boldsymbol{\xi}_5$ are those for cluster B; $\boldsymbol{R}_{\rm AB}$ denotes the relative-motion coordinate between clusters A and B; $\boldsymbol{R}_{\rm cm}$ is the CM coordinate of the two-cluster system; and $\boldsymbol{R}_{\rm A}$ and $\boldsymbol{R}_{\rm B}$ are the CM coordinates of clusters A and B, respectively.

Assuming a Gaussian wave function for each constituent quark, the spatial wave function for cluster A, centered at $\boldsymbol{S}_{\rm A}$, can be written as
\begin{align}
\phi_{\rm A}\!\left(\boldsymbol{r}_1,\boldsymbol{r}_2,\boldsymbol{r}_3; \boldsymbol{S}_{\rm A}\right) = \prod_{i=1}^3 \left(\frac{m_i \omega_{\rm A}}{\pi}\right)^{3/4} {\rm Exp}\left[-\frac{m_i \omega_{\rm A}}{2}\left(\boldsymbol{r}_i - \boldsymbol{S}_{\rm A}\right)^2\right],   \label{eq:wf_B_A}
\end{align}
or, in terms of the Jacobi coordinates defined in Eqs.~\eqref{Jacobi-i}--\eqref{Jacobi-f}, as
\begin{align}
\phi_{\rm A}\!\left(\boldsymbol{r}_1,\boldsymbol{r}_2,\boldsymbol{r}_3; \boldsymbol{S}_{\rm A}\right) = \phi_{\rm A}^{\rm int}\!\left(\boldsymbol{\xi}_1,\boldsymbol{\xi}_2\right) \phi_{\rm A}^{\rm cm}\!\left(\boldsymbol{R}_{\rm A}; \boldsymbol{S}_{\rm A}\right),  \label{eq:wf-spatial-A}
\end{align}
where the internal wave function $\phi_{\rm A}^{\rm int}\!\left(\boldsymbol{\xi}_1,\boldsymbol{\xi}_2\right)$ and the CM-motion wave function $\phi_{\rm A}^{\rm cm}\!\left(\boldsymbol{R}_{\rm A}; \boldsymbol{S}_{\rm A}\right)$ of cluster A are
\begin{align}
\phi_{\rm A}^{\rm int}\!\left(\boldsymbol{\xi}_1,\boldsymbol{\xi}_2\right) &= \prod_{i=1}^2 \left(\frac{M_i\omega_{\rm A}}{\pi}\right)^{3/4} {\rm Exp}\left[-\frac{M_i\omega_{\rm A}}{2} \boldsymbol{\xi}_i^2\right],  \\[6pt]
\phi_{\rm A}^{\rm cm}\!\left(\boldsymbol{R}_{\rm A}; \boldsymbol{S}_{\rm A}\right) &= \left(\frac{M_{\rm A} \omega_{\rm A}}{\pi}\right)^{3/4} {\rm Exp}\left[-\frac{M_{\rm A} \omega_{\rm A}}{2}\left(\boldsymbol{R}_{\rm A} - \boldsymbol{S}_{\rm A}\right)^2\right],
\end{align}
 with
\begin{align}
M_1 &= \frac{m_1 m_2}{m_1+m_2}, \\[6pt]
M_2 &= \frac{\left(m_1+m_2\right)m_3}{m_1+m_2+m_3}.
\end{align}
Similarly, the spatial wave function $\phi_{\rm B}\!\left(\boldsymbol{r}_4,\boldsymbol{r}_5,\boldsymbol{r}_6; \boldsymbol{S}_{\rm B}\right)$ for cluster B, centered at $\boldsymbol{S}_{\rm B}$, can be constructed:
\begin{align}
\phi_{\rm B}\!\left(\boldsymbol{r}_4,\boldsymbol{r}_5,\boldsymbol{r}_6; \boldsymbol{S}_{\rm B}\right) = \prod_{i=4}^6 \left(\frac{m_i \omega_{\rm B}}{\pi}\right)^{3/4} {\rm Exp}\left[-\frac{m_i \omega_{\rm B}}{2}\left(\boldsymbol{r}_i - \boldsymbol{S}_{\rm B}\right)^2\right],
\end{align}
which can be expressed in terms of the Jacobi coordinates as
\begin{align}
\phi_{\rm B}\!\left(\boldsymbol{r}_4,\boldsymbol{r}_5,\boldsymbol{r}_6; \boldsymbol{S}_{\rm B}\right) &= \phi_{\rm B}^{\rm int}\!\left(\boldsymbol{\xi}_4,\boldsymbol{\xi}_5\right) \phi_{\rm B}^{\rm cm}\!\left(\boldsymbol{R}_{\rm B}; \boldsymbol{S}_{\rm B}\right), \label{eq:wf-spatial-B}  
\end{align}
where
\begin{align}
\phi_{\rm B}^{\rm int}\!\left(\boldsymbol{\xi}_4,\boldsymbol{\xi}_5\right) &= \prod_{i=4}^5 \left(\frac{M_i\omega_{\rm B}}{\pi}\right)^{3/4} {\rm Exp}\left[-\frac{M_i\omega_{\rm B}}{2} \boldsymbol{\xi}_i^2\right],  \\[6pt]
\phi_{\rm B}^{\rm cm}\!\left(\boldsymbol{R}_{\rm B}; \boldsymbol{S}_{\rm B}\right) &= \left(\frac{M_{\rm B} \omega_{\rm B}}{\pi}\right)^{3/4} {\rm Exp}\left[-\frac{M_{\rm B} \omega_{\rm B}}{2}\left(\boldsymbol{R}_{\rm B} - \boldsymbol{S}_{\rm B}\right)^2\right],
\end{align}
with
\begin{align}
M_4 &= \frac{m_4 m_5}{m_4+m_5}, \\[6pt]
M_5 &= \frac{\left(m_4+m_5\right)m_6}{m_4+m_5+m_6}.
\end{align}
It should be noted that in previous RGM studies of $BB$ systems at the quark level, the oscillator frequencies for two clusters, $\omega_{\rm A}$ and $\omega_{\rm B}$, were assumed to be identical. Here, following our previous work in Ref.~\cite{Huang:2018}, we treat them as different for general cases. Their specific values are determined by matching the minima of the single-baryon Hamiltonian matrix elements to the experimental masses of the corresponding baryons.

After incorporating the spin, flavor, and color degrees of freedom, the fully antisymmetrized one-baryon wave function is denoted as
\begin{align}
\sum_{f_1,f_2,f_3} &\hat{\phi}_{\rm A}\!\left(\boldsymbol{r}_1,\boldsymbol{r}_2,\boldsymbol{r}_3; \boldsymbol{S}_{\rm A}\right) \nonumber \\[3pt]
&\equiv \sum_{f_1,f_2,f_3} \phi_{\rm A}\!\left(\boldsymbol{r}_1,\boldsymbol{r}_2,\boldsymbol{r}_3; \boldsymbol{S}_{\rm A}\right) \chi^{\rm sfc}_{\rm A}\!\left(f_1,f_2,f_3\right),  \label{eq:wf-3q-A} \\[6pt]
\sum_{f_4,f_5,f_6} &\hat{\phi}_{\rm B}\!\left(\boldsymbol{r}_4,\boldsymbol{r}_5,\boldsymbol{r}_6; \boldsymbol{S}_{\rm B}\right) \nonumber \\[3pt]
&\equiv \sum_{f_4,f_5,f_6} \phi_{\rm B}\!\left(\boldsymbol{r}_4,\boldsymbol{r}_5,\boldsymbol{r}_6; \boldsymbol{S}_{\rm B}\right) \chi^{\rm sfc}_{\rm B}\!\left(f_4,f_5,f_6\right),   \label{eq:wf-3q-B}
\end{align}
where $\chi^{\rm sfc}_{\rm A}\!\left(f_1,f_2,f_3\right)$ is the spin-flavor-color wave function with flavor component $\ket{f_1,f_2,f_3}$ for cluster A, and $\chi^{\rm sfc}_{\rm B}\!\left(f_4,f_5,f_6\right)$ the spin-flavor-color wave function with flavor component $\ket{f_4,f_5,f_6}$ for cluster B.

The primary challenge in constructing the wave function of a two-baryon system with unequal oscillator frequencies lies in the separation of the CM motion of the six-quark system. Following Refs.~\cite{Horiuchi:1977,kamimura:1974}, we introduce a vector $\boldsymbol{S}_{\rm G}$ and the so-called generator coordinate vector $\boldsymbol{S}_i$ defined in terms of $\boldsymbol{S}_{\rm A}$ and $\boldsymbol{S}_{\rm B}$:
\begin{align}
\boldsymbol{S}_{\rm G} &= \frac{M_{\rm AB}}{M_{\rm B}} \boldsymbol{S}_{\rm A} + \frac{M_{\rm AB}}{M_{\rm A}} \boldsymbol{S}_{\rm B},  \\[6pt]
\boldsymbol{S}_i &= \boldsymbol{S}_{\rm A} - \boldsymbol{S}_{\rm B},    \label{eq:Si}
\end{align}
which inversely gives 
\begin{align}
\boldsymbol{S}_{\rm A} &= \boldsymbol{S}_{\rm G} + \frac{M_{\rm AB}}{M_{\rm A}} \boldsymbol{S}_i,  \\[6pt]
\boldsymbol{S}_{\rm B} &= \boldsymbol{S}_{\rm G} - \frac{M_{\rm AB}}{M_{\rm B}} \boldsymbol{S}_i. 
\end{align}
Here $M_{\rm AB} = M_{\rm A} M_{\rm B} / \left(M_{\rm A} + M_{\rm B}\right)$ is the reduced mass of clusters A and B. As usual, we define the following Gaussian functions centered at a series of generator coordinates, $\boldsymbol{S}_i$'s, which serve as a function basis for expanding the relative-motion wave function of two baryon clusters:
\begin{align}
\phi_{\rm rel}\!\left(\boldsymbol{R}_{\rm AB}; \boldsymbol{S}_i\right) = 
\left(\frac{1}{\pi b_{\rm AB}^2}\right)^{3/4} {\rm Exp}\left[-\frac{1}{2b_{\rm AB}^2} \left(\boldsymbol{R}_{\rm AB} - \boldsymbol{S}_i \right)^2\right],  \label{eq:wf-relative}
\end{align}
with
\begin{align}
b_{\rm AB}^2 = \frac{1}{M_{\rm A}\omega_{\rm A}} + \frac{1}{M_{\rm B}\omega_{\rm B}}.
\end{align}
One can then observe that
\begin{align}
& \left(4\pi b_{\rm cm}^2\right)^{-3/4} \int \phi_{\rm A}^{\rm cm}\!\left(\boldsymbol{R}_{\rm A}; \boldsymbol{S}_{\rm G} + \frac{M_{\rm AB}}{M_{\rm A}} \boldsymbol{S}_i\right) \nonumber \\[3pt]
& \qquad\qquad \qquad \times \phi_{\rm B}^{\rm cm}\!\left(\boldsymbol{R}_{\rm B}; \boldsymbol{S}_{\rm G} - \frac{M_{\rm AB}}{M_{\rm B}} \boldsymbol{S}_i\right) {\rm d}\boldsymbol{S}_{\rm G} \nonumber \\[6pt]
& = \phi_{\rm rel}\!\left(\boldsymbol{R}_{\rm AB}; \boldsymbol{S}_i\right),  \label{eq:wf-spatial-SG}
\end{align}
and
\begin{align}
& \left(4\pi b_{\rm cm}^2\right)^{-3/4} \int \phi_{\rm A}^{\rm cm}\!\left(\boldsymbol{R}_{\rm A}; \frac{M_{\rm AB}}{M_{\rm A}} \boldsymbol{S}_i\right) \nonumber \\[3pt]
& \qquad\qquad\qquad \times \phi_{\rm B}^{\rm cm}\!\left(\boldsymbol{R}_{\rm B}; - \frac{M_{\rm AB}}{M_{\rm B}} \boldsymbol{S}_i\right) {\rm d}\boldsymbol{R}_{\rm cm} \nonumber \\[6pt]
& = \phi_{\rm rel}\!\left(\boldsymbol{R}_{\rm AB}; \boldsymbol{S}_i\right),  \label{eq:wf-spatial-Rcm}
\end{align}
where
\begin{align}
b_{\rm cm}^2 = \frac{1}{M_{\rm A}\omega_{\rm A} + M_{\rm B}\omega_{\rm B}}.
\end{align}
Equations \eqref{eq:wf-spatial-SG} and \eqref{eq:wf-spatial-Rcm} are crucial for constructing the total wave function of a two-baryon system. They ensure the wave function is free from spurious CM motion while still being expressible as a product of single Gaussian wave functions for the six quarks. This will be discussed in detail in the following two subsections.

The partial-wave projection of $\phi_{\rm rel}\!\left(\boldsymbol{R}_{\rm AB}; \boldsymbol{S}_i\right)$ defined in Eq.~\eqref{eq:wf-relative} reads
\begin{align}
\int \phi_{\rm rel}\!\left(\boldsymbol{R}_{\rm AB}; \boldsymbol{S}_i\right) Y_{LM_L}\!({\hat{\boldsymbol{S}}_i}) \, {\rm d}{\hat{\boldsymbol{S}}}_i  =  \frac{u^L \!\left(R_{\rm AB}, S_i\right)}{R_{\rm AB}} Y_{LM_L}\!({\hat{\boldsymbol{R}}_{\rm AB}}),
\end{align}
where
\begin{align}
u^L \!\left(R_{\rm AB}, S_i\right) =&\; 4\pi R_{\rm AB} \left(\frac{1}{\pi b_{\rm AB}^2}\right)^{3/4} {\rm Exp}\left[-\frac{1}{2b_{\rm AB}^2} \left( R^2_{\rm AB} + S^2_i \right) \right]  \nonumber \\[3pt]
& \times \; i_L\!\left(R_{\rm AB} S_i / b_{\rm AB}^2 \right),  \label{eq:uL}
\end{align}
with $i_L$ being the $L$-th modified spherical Bessel function, e.g., $i_0(x) = {\sinh(x)}/x$, $i_1(x) = {\cosh(x)}/x - {\sinh(x)}/x^2$.

Generally, the spatial wave function for the relative motion of clusters A and B, denoted as $\chi_{\rm rel}\!\left(\boldsymbol{R}_{\rm AB}\right)$, can be projected into partial waves as
\begin{align}
\chi_{\rm rel}\!\left(\boldsymbol{R}_{\rm AB}\right)  =  \sum_L \frac{\chi^L_{\rm rel} \!\left(R_{\rm AB}\right)}{R_{\rm AB}} Y_{LM_L}\!({\hat{\boldsymbol{R}}_{\rm AB}}).
\end{align}
In the following two subsections, we show in detail how to construct the projected wave function $\chi^L_{\rm rel} \!\left(R_{\rm AB}\right)$ by use of $u^L\!\left(R_{\rm AB}, S_i\right)$ of Eq.~\eqref{eq:uL}, and how to solve it according to the dynamics of the $BB$ system, in cases of bound-state and scattering problems, respectively.

\subsection{Bound state problem}

For a bound-state problem, the projected relative-motion wave function $\chi^L_{\rm rel} \!\left(R_{\rm AB}\right)$ can be expanded as
\begin{align}
\chi^L_{\rm rel} \!\left(R_{\rm AB}\right) = \sum_{i=1}^n c^L_i \, u^L \!\left(R_{\rm AB}, S_i\right), 
\end{align}
which is equivalent to
\begin{align}
\chi_{\rm rel}\!\left(\boldsymbol{R}_{\rm AB}\right) = \sum_{L,i} c^L_i \int \phi_{\rm rel}\!\left(\boldsymbol{R}_{\rm AB}; \boldsymbol{S}_i\right) Y_{LM_L}\!({\hat{\boldsymbol{S}}_i}) \, {\rm d}{\hat{\boldsymbol{S}}}_i,  \label{eq:wf-spatial-relative}
\end{align}
where $n$ is the number of generator coordinates, and $c^L_i$'s are coefficients to be determined by the dynamics of the system. 

The spatial wave function of a $BB$ system can be, in general, constructed as
\begin{align}
\Phi_{\rm AB}\!\left(\boldsymbol{\xi}_1,\boldsymbol{\xi}_2,\boldsymbol{\xi}_4,\boldsymbol{\xi}_5,\boldsymbol{R}_{\rm AB}\right) = \phi_{\rm A}^{\rm int}\!\left(\boldsymbol{\xi}_1,\boldsymbol{\xi}_2\right) \phi_{\rm B}^{\rm int}\!\left(\boldsymbol{\xi}_4,\boldsymbol{\xi}_5\right) \chi_{\rm rel}\!\left(\boldsymbol{R}_{\rm AB}\right).
\end{align}
By use of Eqs.~\eqref{eq:wf-spatial-A}, \eqref{eq:wf-spatial-B}, \eqref{eq:wf-spatial-SG}, \eqref{eq:wf-spatial-Rcm}, and \eqref{eq:wf-spatial-relative}, it can be expressed either as
\begin{align}
&\Phi_{\rm AB}\!\left(\boldsymbol{\xi}_1,\boldsymbol{\xi}_2,\boldsymbol{\xi}_4,\boldsymbol{\xi}_5,\boldsymbol{R}_{\rm AB}\right)  \nonumber \\[6pt]
& = \left(4\pi b_{\rm cm}^2\right)^{-3/4} \sum_{L,i} c^L_i \int \phi_{\rm A}\!\left(\boldsymbol{r}_1,\boldsymbol{r}_2,\boldsymbol{r}_3; \boldsymbol{S}_{\rm G} + \frac{M_{\rm AB}}{M_{\rm A}} \boldsymbol{S}_i\right) \nonumber \\[3pt]
& \quad \times \phi_{\rm B}\!\left(\boldsymbol{r}_4,\boldsymbol{r}_5,\boldsymbol{r}_6; \boldsymbol{S}_{\rm G} - \frac{M_{\rm AB}}{M_{\rm B}} \boldsymbol{S}_i\right) Y_{LM_L}\!({\hat{\boldsymbol{S}}_i}) \, {\rm d}{\hat{\boldsymbol{S}}}_i \, {\rm d}\boldsymbol{S}_{\rm G},  \label{eq:wf-r-BB-SG}
\end{align}
or as
\begin{align}
&\Phi_{\rm AB}\!\left(\boldsymbol{\xi}_1,\boldsymbol{\xi}_2,\boldsymbol{\xi}_4,\boldsymbol{\xi}_5,\boldsymbol{R}_{\rm AB}\right)   \nonumber \\[6pt]
& = \left(4\pi b_{\rm cm}^2\right)^{-3/4} \sum_{L,i} c^L_i \int \phi_{\rm A}\!\left(\boldsymbol{r}_1,\boldsymbol{r}_2,\boldsymbol{r}_3; \frac{M_{\rm AB}}{M_{\rm A}} \boldsymbol{S}_i\right) \nonumber \\[3pt]
& \quad \times \phi_{\rm B}\!\left(\boldsymbol{r}_4,\boldsymbol{r}_5,\boldsymbol{r}_6; - \frac{M_{\rm AB}}{M_{\rm B}} \boldsymbol{S}_i\right) Y_{LM_L}\!({\hat{\boldsymbol{S}}_i}) \, {\rm d}{\hat{\boldsymbol{S}}}_i \, {\rm d}\boldsymbol{R}_{\rm cm},   \label{eq:wf-r-BB-Rcm}
\end{align}
One sees that in either case the spatial wave function of the two-baryon system can be expressed in terms of a product of the single-quark wave functions, plus integrations over $\boldsymbol{S}_{\rm G}$ (or $\boldsymbol{R}_{\rm cm}$) and ${\hat{\boldsymbol{S}}}_i$ and summations over $i$ and $L$. Expressing the spatial wave function of the two-baryon system as a product of single-quark wave functions is crucial in practice for calculating the normalization and interaction matrix elements. 

After incorporating the spin, flavor, and color degrees of freedom, by use of the Eqs.~\eqref{eq:wf-3q-A}, \eqref{eq:wf-3q-B}, \eqref{eq:wf-r-BB-SG}, and \eqref{eq:wf-r-BB-Rcm}, the fully antisymmetrized two-baryon wave function can be denoted as either
\begin{align}
&\sum_f \hat{\Phi}_{\rm AB}\!\left(\boldsymbol{\xi}_1,\boldsymbol{\xi}_2,\boldsymbol{\xi}_4,\boldsymbol{\xi}_5,\boldsymbol{R}_{\rm AB}\right) \nonumber \\[3pt]
& = \sum_{L,i,f} c^L_i \left(4\pi b_{\rm cm}^2\right)^{-3/4} \int {\rm d}\boldsymbol{S}_{\rm G} \, {\rm d}{\hat{\boldsymbol{S}}}_i \, Y_{LM_L}\!({\hat{\boldsymbol{S}}_i})  \nonumber \\[3pt]
& \qquad \times {\cal A}\left[ \hat{\phi}_{\rm A}\!\left(\boldsymbol{r}_1,\boldsymbol{r}_2,\boldsymbol{r}_3; \boldsymbol{S}_{\rm G} + \frac{M_{\rm AB}}{M_{\rm A}} \boldsymbol{S}_i\right) \right. \nonumber \\[3pt]
& \quad ~~\qquad \times \left.\hat{\phi}_{\rm B}\!\left(\boldsymbol{r}_4,\boldsymbol{r}_5,\boldsymbol{r}_6; \boldsymbol{S}_{\rm G} - \frac{M_{\rm AB}}{M_{\rm B}} \boldsymbol{S}_i\right) \right]_{SYT},  \label{eq:wf_BB-SG}
\end{align}
or
\begin{align}
&\sum_f \hat{\Phi}_{\rm AB}\!\left(\boldsymbol{\xi}_1,\boldsymbol{\xi}_2,\boldsymbol{\xi}_4,\boldsymbol{\xi}_5,\boldsymbol{R}_{\rm AB}\right) \nonumber \\[3pt]
& = \sum_{L,i,f} c^L_i \left(4\pi b_{\rm cm}^2\right)^{-3/4} \int {\rm d}\boldsymbol{R}_{\rm cm} \, {\rm d}{\hat{\boldsymbol{S}}}_i \, Y_{LM_L}\!({\hat{\boldsymbol{S}}_i}) \nonumber \\[3pt]
& \qquad \times  {\cal A}\left[ \hat{\phi}_{\rm A}\!\left(\boldsymbol{r}_1,\boldsymbol{r}_2,\boldsymbol{r}_3; \frac{M_{\rm AB}}{M_{\rm A}} \boldsymbol{S}_i\right) \right. \nonumber \\[3pt]
& \quad ~~\qquad \times \left.\hat{\phi}_{\rm B}\!\left(\boldsymbol{r}_4,\boldsymbol{r}_5,\boldsymbol{r}_6; - \frac{M_{\rm AB}}{M_{\rm B}} \boldsymbol{S}_i\right) \right]_{SYT}, \label{eq:wf_BB-Rcm}
\end{align}
where $f$ is an abbreviation of $f_1,f_2,f_3,f_4,f_5$, and $f_6$; $S$, $Y$, and $T$ are the spin, hyper charge, and isospin of the two-baryon system, respectively; and $\cal A$ is the antisymmetrizer for quarks between two clusters A and B,
\begin{align}
{\cal A} = \left(1-9P_{36}\right)\left(1-P_{\rm AB}\right),  \label{eq:anti-symmtrizer}
\end{align}
with $P_{36}$ being the exchange operator between the 3rd and 6th quarks, and $P_{\rm AB}$ the exchange operator between clusters A and B. 

The wave function $\sum_f \hat{\Phi}_{\rm AB}\!\left(\boldsymbol{\xi}_1,\boldsymbol{\xi}_2,\boldsymbol{\xi}_4,\boldsymbol{\xi}_5,\boldsymbol{R}_{\rm AB}\right)$ satisfies the variational formulation of the Schr\"odinger equation:
\begin{align}
\sum_{f',f}\braket{\delta\hat{\Phi}_{\rm A'B'} | H - E | \hat{\Phi}_{\rm AB}} = 0,   \label{eq:RGM-bound-1}
\end{align} 
where the integration is performed over the two-baryon system's internal coordinates $\boldsymbol{\xi}_1$, $\boldsymbol{\xi}_2$, $\boldsymbol{\xi}_4$, $\boldsymbol{\xi}_5$, and $\boldsymbol{R}_{\rm AB}$. More explicitly, this equation can be expressed as
\begin{align}
\sum_{L,i} \left( {\cal H}^{L'L}_{ji} - E {\cal N}^{L'L}_{ji} \right) c^L_i = 0,  \label{eq:RGM-bound-2}
\end{align}
with
\begin{widetext}
\begin{align}
\left\{ \begin{array}{c} {\cal H}^{L'L}_{ji} \\ {\cal N}^{L'L}_{ji} \end{array} \right\} \equiv & \sum_{f',f} \left(4\pi {b'}_{\!\!\rm cm}^2\right)^{-3/4} \left(4\pi b_{\rm cm}^2\right)^{-3/4} \int \left[ \hat{\phi}^\dag_{\rm A'}\!\left(\boldsymbol{r}_1,\boldsymbol{r}_2,\boldsymbol{r}_3; \boldsymbol{S}_{\rm G} + \frac{M_{\rm A'B'}}{M_{\rm A'}} \boldsymbol{S}_j\right) \hat{\phi}^\dag_{\rm B'}\!\left(\boldsymbol{r}_4,\boldsymbol{r}_5,\boldsymbol{r}_6; \boldsymbol{S}_{\rm G} - \frac{M_{\rm A'B'}}{M_{\rm B'}} \boldsymbol{S}_j\right) \right]_{S'YT} \left\{ \begin{array}{c} H \\ 1 \end{array} \right\} \nonumber \\[3pt]
& \times  {\cal A} \left[ \hat{\phi}_{\rm A}\!\left(\boldsymbol{r}_1,\boldsymbol{r}_2,\boldsymbol{r}_3; \frac{M_{\rm AB}}{M_{\rm A}} \boldsymbol{S}_i\right) \hat{\phi}_{\rm B}\!\left(\boldsymbol{r}_4,\boldsymbol{r}_5,\boldsymbol{r}_6; -\frac{M_{\rm AB}}{M_{\rm B}} \boldsymbol{S}_i\right) \right]_{SYT} Y^*_{L'M_{L'}}\!({\hat{\boldsymbol{S}}_j}) \, Y_{LM_L}\!({\hat{\boldsymbol{S}}_i}) \, {\rm d}{\boldsymbol{r}}_1 {\rm d}{\boldsymbol{r}}_2 {\rm d}{\boldsymbol{r}}_3 {\rm d}{\boldsymbol{r}}_4 {\rm d}{\boldsymbol{r}}_5 {\rm d}{\boldsymbol{r}}_6 {\rm d}\boldsymbol{S}_{\rm G} {\rm d}{\hat{\boldsymbol{S}}}_j  {\rm d}{\hat{\boldsymbol{S}}}_i.   \label{eq:RGM-Lji}
\end{align}
\end{widetext}
To obtain the expression of Eq.~\eqref{eq:RGM-Lji}, the wave function $\sum_f \hat{\Phi}_{\rm AB}$---whose spatial variables are two-baryon internal coordinates---is rewritten, by use of Eqs.~\eqref{eq:wf_BB-SG} and \eqref{eq:wf_BB-Rcm}, as a product of single-quark wave functions. This procedure is essential for the practical calculation of normalization and interaction matrix elements. In the expression of Eq.~\eqref{eq:RGM-Lji}, the integrations over ${\hat{\boldsymbol{S}}}_i$ and ${\hat{\boldsymbol{S}}}_j$ are performed for partial-wave projection, and the integration over $\boldsymbol{S}_{\rm G}$ is performed to ensure that the product of Jacobi coordinates can be swapped with that of single-quark coordinates, i.e. ${\rm d}{\boldsymbol{\xi}}_1 {\rm d}{\boldsymbol{\xi}}_2 {\rm d}{\boldsymbol{\xi}}_4 {\rm d}{\boldsymbol{\xi}}_5 {\rm d}{\boldsymbol{R}}_{\rm AB} {\rm d}{\boldsymbol{R}}_{\rm cm} = {\rm d}{\boldsymbol{r}}_1 {\rm d}{\boldsymbol{r}}_2 {\rm d}{\boldsymbol{r}}_3 {\rm d}{\boldsymbol{r}}_4 {\rm d}{\boldsymbol{r}}_5 {\rm d}{\boldsymbol{r}}_6$.

Equation \eqref{eq:RGM-bound-2} holds for any orbital angular momentum $L'$ and generator coordinate ${\boldsymbol{S}}_j$ ($j=1,2,3,\cdots,n$). Physically, of course, $L'$ is restricted by the parity and total angular momentum of the considered $BB$ system. Solving this set of RGM equations, one can get the coefficients $c^L_i$'s for the two-baryon wave function and the binding energy $E$ of the $BB$ system.

\subsection{Scattering problem}

For a scattering problem, the projected relative-motion wave function $\chi^L_{\rm rel} \!\left(R_{\rm AB}\right)$ can be expanded as
\begin{align}
\chi^L_{\rm rel} \!\left(R_{\rm AB}\right) = \sum_{i=1}^n c^L_i \, {\tilde u}^L \!\left(R_{\rm AB}, S_i\right),  \label{eq:wf-scattering-0}
\end{align}
with
\begin{align}
& {\tilde u}^L \!\left(R_{\rm AB}, S_i\right) \nonumber \\[6pt]
& \quad \equiv \left\{ \begin{array}{lcl} p_i^L u^L \!\left(R_{\rm AB}, S_i\right),  &&  R_{\rm AB} \le R_c \\[6pt] \left[ h_L^-\!\left(k_{\rm AB} R_{\rm AB}\right) - s_i^L h_L^+\!\left(k_{\rm AB} R_{\rm AB}\right) \right] R_{\rm AB}, && R_{\rm AB} \ge R_c \end{array} \right.   \label{eq:wf-scattering}
\end{align}
where $h_L^\pm$ are the $L$-th spherical Hankel functions, e.g., $h_0^\pm(x)=e^{\pm\,ix}/x$, $h_1^\pm(x)=e^{\pm\,ix}/x^2\mp ie^{\pm\,ix}/x$; $k_{\rm AB} = \sqrt{2M_{\rm AB} E_{\rm rel}}$ is the magnitude of three-momentum for the two-baryon relative motion, with $E_{\rm rel} = E - E_{\rm A} - E_{\rm B}$ being the relative energy between two clusters A and B ($E$, $E_{\rm A}$, and $E_{\rm B}$ are the total energy of the $BB$ system and the internal energies of clusters A and B, respectively); $R_c$ is the cut-off radius beyond which all strong interactions between the two clusters can be neglected; and $p_i^L$ and $s_i^L$ are complex parameters to be determined by the smoothness condition of ${\tilde u}^L \!\left(R_{\rm AB}, S_i\right)$ at $R_{\rm AB} = R_c$, which results in
\begin{align}
\left\{\begin{array}{l} p_i^L = \dfrac{ b_{\rm AB}^2 \, z \left[h_{L+1}^+(z) h_L^-(z) - h_L^+(z) h_{L+1}^-(z) \right] }{ k_{\rm AB} b_{\rm AB}^2 u^L_i h_{L+1}^+(z) - R_c u^L_i h_L^+(z) + S_i \, u^{L+1}_i h_L^+(z) },  \\[15pt]  s_i^L= \dfrac{ k_{\rm AB} b_{\rm AB}^2 u^L_i h_{L+1}^-(z) - R_c u^L_i h_L^-(z) + S_i \, u^{L+1}_i h_L^-(z) }{ k_{\rm AB} b_{\rm AB}^2 u^L_i h_{L+1}^+(z) - R_c u^L_i h_L^+(z) + S_i \, u^{L+1}_i h_L^+(z) },  \end{array}  \right.
\end{align}
with $z \equiv k_{\rm AB} R_c$ and $u^L_i\equiv u^L \!\left(R_c, S_i\right)$.

Generally, in the region $R_{\rm AB} \ge R_c$ where the strong interaction between the two baryons can be neglected, the projected relative-motion wave function $\chi^L_{\rm rel} \!\left(R_{\rm AB}\right)$ should take the asymptotic form:
\begin{align}
\chi^L_{\rm rel} \!\left(R_{\rm AB}\right) = \left[ h_L^-\!\left(k_{\rm AB} R_{\rm AB}\right) - S_L h_L^+\!\left(k_{\rm AB} R_{\rm AB}\right) \right] R_{\rm AB},  \label{eq:wf-asymptoic}
\end{align}
where $S_L$ is the S-matrix element, and in the single-channel case, 
\begin{equation}
S_L = e^{2i\delta_L}.
\end{equation}
Comparing Eq.~\eqref{eq:wf-asymptoic} with Eq.~\eqref{eq:wf-scattering-0}, we obtain
\begin{align}
\sum_{i=1}^n c^L_i &= 1,   \label{eq:ci-condition}  \\[3pt]
\sum_{i=1}^n c^L_i s_i^L &= S_L.  \label{eq:ci-SL}
\end{align}

Due to Eq.~\eqref{eq:ci-condition}, the projected relative-motion wave function  $\chi^L_{\rm rel} \!\left(R_{\rm AB}\right)$ in the region $R_{\rm AB} \le R_c$ can be expressed as
\begin{align}
\chi^L_{\rm rel} \!\left(R_{\rm AB}\right) = & \; \sum_{i=1}^n c^L_i  p^L_i\, u^L \!\left(R_{\rm AB}, S_i\right)  \nonumber \\[3pt]
= & \; \sum_{i=1}^{n-1} c^L_i \left[ p_i^L\, u^L \!\left(R_{\rm AB}, S_i\right) - p_n^L\, u^L \!\left(R_{\rm AB}, S_n\right) \right] \nonumber \\[3pt]
& \; + p_n^L\, u^L \!\left(R_{\rm AB}, S_n\right).  \label{RGM-scatter-rel-wf}
\end{align}
The coefficients $c^L_i$ can, in principle, be obtained by solving Eq.~\eqref{eq:RGM-bound-1}. However, note that the integration over the relative-motion coordinate $R_{\rm AB}$ is restricted to the interval $[0, R_c]$, which prevents a direct transformation of the integration variables from Jacobi coordinates to single-quark coordinates. This issue can be resolved by subtracting the integration over $R_{\rm AB}$ from $R_c$ to $\infty$. Note that in the region $R_{\rm AB} \ge R_c$, there is no interaction between the clusters, and only the relative kinetic energy remains. Then, the matrix elements of $\left(H-E\right)$ within the range $0 \le R_{\rm AB} \le R_c$, denoted by ${\cal K}^{L'L}_{ji}$, can be expressed as
\begin{align}
{\cal K}^{L'L}_{ji} = p_j^{L'} p_i^L \left( {\cal H}_{ji}^{L'L} - E {\cal N}_{ji}^{L'L} \right) - \delta_{L'L} {{\cal X}}_{ji}^L,
\end{align} 
with ${\cal H}_{ji}^{L'L}$ and ${\cal N}_{ji}^{L'L}$ being defined in Eq.~\eqref{eq:RGM-Lji}, and ${{\cal X}}_{ji}^L$ being the matrix elements of $\left(H-E\right)$ for $R_{\rm AB} \ge R_c$. In practice, ${{\cal X}}_{ji}^L$ can be calculated as
\begin{align}
{{\cal X}}_{ji}^L = & \; p_j^L p_i^L \int_{R_c}^\infty u^L\!\left(R_{\rm AB}, S_j\right) \left[ -\frac{1}{2M_{\rm AB}} \frac{{\rm d}^2}{{\rm d}R_{\rm AB}^2} \right. \nonumber \\[3pt]
& \; + \left. \frac{1}{2M_{\rm AB}} \frac{L\left(L+1\right)}{R_{\rm AB}^2} - E_{\rm rel} \right]  u^L\!\left(R_{\rm AB}, S_i\right) {\rm d}R_{\rm AB}.
\end{align}

The two-baryon wave function constructed by use of the projected relative-motion wave function of Eq.~\eqref{RGM-scatter-rel-wf} satisfies the variational formulation of the Schr\"odinger equation,  Eq.~\eqref{eq:RGM-bound-1}, which directly results in
\begin{align}
\sum_L \left( \sum_{i=1}^{n-1} {\cal L}_{ji}^{L'L} c^L_i - {\cal M}^{L'L}_j \right) = 0,  \label{eq:RGM-scatter}
\end{align}
with
\begin{align}
{\cal L}^{L'L}_{ji} &= {\cal K}^{L'L}_{ji} - {\cal K}^{L'L}_{ni} - {\cal K}^{L'L}_{jn} + {\cal K}^{L'L}_{nn}, \\[6pt]
{\cal M}^{L'L}_j &= {\cal K}^{L'L}_{nn} - {\cal K}^{L'L}_{jn}.
\end{align}

Again, Eq.~\eqref{eq:RGM-scatter} holds for any $L'$ and ${\boldsymbol{S}}_j$ ($j=1,2,3,\cdots,n-1$). Solving this set of RGM equations, one can get the coefficients $c^L_i$'s for the two-baryon scattering wave function and consequently the S-matrix elements by use of Eq.~\eqref{eq:ci-SL}.

\section{The Chiral SU(3) quark model}  \label{Sec:chiral-QM}

Although the constituent quark model incorporating OGE and a confinement potential has achieved considerable success in describing the static properties of single hadrons, as is well known, it suffers from significant deficiencies in at least two aspects. First, it lacks essential medium- and long-range attraction in the $NN$ interaction. Second, it fails to provide a satisfactory explanation for the origin of the large constituent quark mass,  which is salient compared to the tiny current quark mass. These shortcomings can be effectively addressed within the framework of the chiral quark model. In this model, the coupling between quarks and chiral fields is introduced, yielding a quark-quark interaction Lagrangian that is manifestly invariant under chiral symmetry transformations. Here, the valence quarks acquire their dynamical (constituent) masses through  spontaneous chiral symmetry breaking, while the Goldstone bosons obtain their physical masses via explicit chiral symmetry breaking induced by the small current quark masses. Crucially, the one-boson exchange (OBE) mechanism provides the necessary medium- and long-range attraction for the $NN$ interaction. Consequently, the chiral quark model is regarded as a well-motivated framework capable of describing both  hadron spectroscopy and hadron-hadron interactions. 

In the present work, we employ the chiral SU(3) quark model, whose details can be found in Refs.~\cite{Zhang:1997,Huang:2004,Huang:2004-2,Huang:2018}. Below, we briefly outline its main features.

The interaction Lagrangian coupling quarks to chiral fields in the flavor SU(3) case can be obtained by a straightforward extension of the quark-level SU(2) linear $\sigma$-model, which gives
\begin{equation}  \label{eq:L_ch}
{\cal L}_I^{\rm ch} = - g_{\rm ch} \bar{\psi} \left[ \sum^{8}_{a=0} \left( \sigma_a  + i \gamma_5 \pi_a  \right) \lambda^a \right] \psi.
\end{equation}
Here, $\psi$ represents the quark field; $\pi_a$ and $\sigma_a$ $(a=0,1,...,8)$ denote the nonet pseudoscalar and scalar fields, respectively; $\lambda^{a}$ are the Gell-Mann matrices of the flavor SU(3) group; and $g_{\rm ch}$ is the quark and chiral field coupling constant. The Lagrangian in Eq.~\eqref{eq:L_ch} is invariant under infinitesimal chiral SU(3)$_L\,\times\,$SU(3)$_R$ transformations. 

In practice, a form factor $F(\boldsymbol{q}^{2})$ is introduced at the vertices of quark and chiral field coupling, which is supposed to effectively account for the internal structure of the chiral fields: 
\begin{equation}  \label{eq:FF}
F(\boldsymbol{q}^{2})=\left( \frac{\Lambda^2}{\Lambda^2 + \boldsymbol{q}^2}\right)^{1/2},
\end{equation}
where $\boldsymbol{q}$ is the three momentum of the exchanged boson, and $\Lambda$ is a cutoff mass indicating the chiral symmetry breaking scale \cite{kusa:1991,buchmann:1991,henley:1990}. 

From Eqs.~\eqref{eq:L_ch} and \eqref{eq:FF}, one derives the following potentials between the $i$th and $j$th constituent quarks, induced by quark and chiral field couplings:
\begin{align}
V^{\rm ch}_{ij} = \sum_{a=0}^8 \left( V^{\sigma_a}_{ij} + V^{\pi_a}_{ij} \right),   \label{eq:V_ch} 
\end{align}
with
\begin{align}
V^{\sigma_a}_{ij} =&~ V^{\sigma_a}_{\rm cen}(\boldsymbol{r}_{ij}) + V^{\sigma_a}_{\rm ls}(\boldsymbol{r}_{ij}),  \label{eq:V_q-q-sigma}  \\[6pt]
V^{\sigma_a}_{\rm cen}(\boldsymbol{r}_{ij}) =& - C(g_{\rm ch}, m'_{\sigma_a}, \Lambda) Y_1(m'_{\sigma_a}, \Lambda, r_{ij}) \left(\lambda^a_i \lambda^a_j\right),    \\[6pt]
V^{\sigma_a}_{\rm ls}(\boldsymbol{r}_{ij}) = & - C(g_{\rm ch},m'_{\sigma_a}, \Lambda) \frac{m'^{\,2}_{\sigma_a} s^a_{ij}}{8} Z_3(m'_{\sigma_a}, \Lambda, r_{ij})    \nonumber \\[3pt]
&\, \times \left(\lambda^a_i \lambda^a_j\right) \left[\boldsymbol{L} \cdot \left(\boldsymbol{\sigma}_i + \boldsymbol{\sigma}_j\right)\right],  \\[6pt]
V^{\pi_a}_{ij} =&~ V^{\pi_a}_{\rm cen}(\boldsymbol{r}_{ij}) + V^{\pi_a}_{\rm ten}(\boldsymbol{r}_{ij}),  \label{eq:V_q-q-pi}   \\[6pt]
V^{\pi_a}_{\rm cen}(\boldsymbol{r}_{ij}) =&~ C(g_{\rm ch}, m'_{\pi_a}, \Lambda) \frac{m'^{\,2}_{\pi_a} c^a_{ij}}{48} Y_3(m'_{\pi_a}, \Lambda, r_{ij})  \nonumber \\[3pt]
& \, \times \left(\boldsymbol{\sigma}_i \cdot \boldsymbol{\sigma}_j \, \lambda^a_i \lambda^a_j\right),   \\[6pt]
V^{\pi_a}_{\rm ten}(\boldsymbol{r}_{ij}) = &~  C(g_{\rm ch}, m'_{\pi_a}, \Lambda) \frac{m'^{\,2}_{\pi_a} c^a_{ij}}{48} H_3(m'_{\pi_a}, \Lambda, r_{ij})\left(\lambda^a_i \lambda^a_j\right)  \nonumber\\[3pt]
&\, \times \left[3\left(\boldsymbol{\sigma}_i \cdot \hat{\boldsymbol{r}}_{ij}\right)\left(\boldsymbol{\sigma}_j \cdot \hat{\boldsymbol{r}}_{ij}\right)- \boldsymbol{\sigma}_i \cdot \boldsymbol{\sigma}_j\right],
\end{align}
where the subscripts ``cen", ``ls", and ``ten" denote central, spin-orbit, and tensor forces, respectively, and 
\begin{align}
C(g_{\rm ch},m,\Lambda) &= \frac{g^2_{\rm ch}}{4\pi} \frac{\Lambda^2}{\Lambda^2-m^2} m,  \\[5pt]
Y_1(m,\Lambda,r) &= Y(mr) - \frac{\Lambda}{m} Y(\Lambda r),  \label{x1mlr}  \\[5pt]
Y_3(m,\Lambda,r) &= Y(mr) - \left(\frac{\Lambda}{m}\right)^3 Y(\Lambda r),  \\[5pt]
Z_3(m,\Lambda,r) &= Z(mr) - \left(\frac{\Lambda}{m}\right)^3 Z(\Lambda r),  \\[5pt]
H_3(m,\Lambda,r) &= H(mr) - \left(\frac{\Lambda}{m}\right)^3 H(\Lambda r),  \\[5pt]
Y(x) &= \frac{1}{x}e^{-x},  \\[5pt]
Z(x) &= \left(\frac{1}{x}+\frac{1}{x^2}\right)Y(x),  \\[5pt]
H(x) &= \left(1+\frac{3}{x}+\frac{3}{x^2}\right)Y(x),  \\[5pt]
c^a_{ij} &= \left\{\begin{array}{lcl} \dfrac{4}{m_i m_j},  && (a=0,1,2,3,8) \\[9pt]  \dfrac{\left(m_i + m_j\right)^2}{m_i^2 m_j^2},  && (a=4,5,6,7) \end{array} \right.   \\[5pt]
s^a_{ij} &= \left\{\begin{array}{lccl} \dfrac{1}{m_i^2}+\dfrac{1}{m_j^2},  &&& \,~(a=0,1,2,3,8) \\[12pt]   \dfrac{2}{m_i m_j}.   &&& \,~(a=4,5,6,7) \end{array} \right.
\end{align}
Here, $m_i$ and $m_j$ are the masses of the $i$th and $j$th constituent quarks, and $m'_{\sigma_a}$ and $m'_{\pi_a}$ are related to the scalar meson mass $m_{\sigma_a}$ and pseudoscalar meson mass $m_{\pi_a}$, respectively, by
\begin{align}  \label{eq:mass_S}
m_{\sigma_a}' &= \left\{\begin{array}{lcl} m_{\sigma_a},   && (a=0,1,2,3,8) \\[5pt]    \sqrt{m_{\sigma_a}^2-\left(m_i-m_j\right)^2},   && (a=4,5,6,7) \end{array} \right.  \\[5pt]
m_{\pi_a}' &= \left\{\begin{array}{lcl} m_{\pi_a},   && \,(a=0,1,2,3,8) \\[5pt]    \sqrt{m_{\pi_a}^2-\left(m_i-m_j\right)^2}.   && \,(a=4,5,6,7) \end{array} \right.  \label{eq:mass_PS}
\end{align}
The relations for $a=4,5,6,7$ in Eqs.~\eqref{eq:mass_S} and \eqref{eq:mass_PS} arise from an explicit treatment of the mass difference between $u(d)$ and $s$ quarks when deriving the potentials in Eqs.~\eqref{eq:V_q-q-sigma} and \eqref{eq:V_q-q-pi} for $\kappa$- and $K$-meson exchanges from the quark and chiral field interaction Lagrangian in Eq.~\eqref{eq:L_ch}.
    
For pseudoscalar meson exchanges, the physical $\eta$ and $\eta$' are treated as mixtures of the SU(3) eigenstates $\eta_0$ and $\eta_8$:
\begin{equation}
\left\{ \begin{array}{l} \eta = \eta_8 \cos\theta^{PS} - \eta_0 \sin\theta^{PS}, \\[6pt]  \eta' = \eta_8 \sin\theta^{PS} + \eta_0 \cos\theta^{PS}, \end{array} \right.
\end{equation}
with the mixing angle $\theta^{PS}$ set to the empirical value $\theta^{PS} = -23^\circ$.

In addition to the OBE potentials in Eq.~\eqref{eq:V_ch}, the description of hadron structures and hadron-hadron dynamics requires the inclusion of the OGE potential, which accounts for the short-range perturbative effects,
\begin{align}
V^{\rm OGE}_{ij}  = V^{\rm OGE}_{\rm cen}({\boldsymbol r}_{ij}) + V^{\rm OGE}_{\rm ls}({\boldsymbol r}_{ij})  + V^{\rm OGE}_{\rm ten}({\boldsymbol r}_{ij}),
\end{align}
with
\begin{align}
V^{\rm OGE}_{\rm cen}({\boldsymbol r}_{ij})  = &~ \frac{g_i g_j}{4} \left(\boldsymbol{\lambda}^c_i \cdot \boldsymbol{\lambda}^c_j\right) \left\{\frac{1}{r_{ij}}-\frac{\pi}{2} \delta(\boldsymbol{r}_{ij}) \left[ \frac{1}{m^2_i}+\frac{1}{m^2_j} \right.\right. \nonumber \\[3pt]
&~ \left. + \left. \frac{4}{3}\frac{1}{m_i m_j} \left( \boldsymbol{\sigma}_i \cdot \boldsymbol{\sigma}_j \right) \right] \right\},   \\[6pt]
V^{\rm OGE}_{\rm ls}({\boldsymbol r}_{ij}) = &~ -\frac{g_i g_j}{4} \left(\boldsymbol{\lambda}^c_i \cdot \boldsymbol{\lambda}^c_j\right) \frac{m_i^2+m_j^2+4m_im_j}{8m_i^2 m_j^2}\frac{1}{r^3_{ij}}  \nonumber \\[3pt]
&~ \times  \left[ {\boldsymbol{L} \cdot \left( \boldsymbol{\sigma}_i + \boldsymbol{\sigma}_j \right)} \right],  \\[6pt]
V^{\rm OGE}_{\rm ten}({\boldsymbol r}_{ij}) = &~ -\frac{g_ig_j}{4} \left(\boldsymbol{\lambda}^c_i \cdot \boldsymbol{\lambda}^c_j\right) \frac{1}{4m_i m_j}\frac{1}{r^3_{ij}}  \nonumber \\[3pt]
&~ \times \left[3\left(\boldsymbol{\sigma}_i \cdot \hat{\boldsymbol{r}}_{ij}\right)\left(\boldsymbol{\sigma}_j \cdot \hat{\boldsymbol{r}}_{ij}\right)- \boldsymbol{\sigma}_i \cdot \boldsymbol{\sigma}_j\right],
\end{align}
and a phenomenological confinement potential describing the long-range non-perturbative effects, usually taken to be of quadratic or linear form,
\begin{equation}
V_{ij}^{\rm conf} = \left\{\begin{array}{lcl}  -\big(\boldsymbol{\lambda}_{i}^{c}\cdot\boldsymbol{\lambda}_{j}^{c}\big) \big(a_{ij}r_{ij}^2 +a_{ij}^{0}\big),  && {\rm (Model~I)}  \\[8pt] -\big(\boldsymbol{\lambda}_{i}^{c}\cdot\boldsymbol{\lambda}_{j}^{c}\big) \big(a_{ij}r_{ij} +a_{ij}^{0}\big).  && {\rm (Model~II)}
\end{array}\right.   \label{eq:conf}
\end{equation}
The OGE coupling constants $g_i$ and $g_j$, the confinement strength $a_{ij}$, and the zero-point energy $a_{ij}^{0}$ are model parameters, whose determination will be discussed in the next subsection.

Finally, the total Hamiltonian of the chiral SU(3) quark model is given by
\begin{align}
H = \sum_{i=1}^{n_q} \left(m_i + \frac{\boldsymbol{p}_i^2}{2m_i} \right) - \frac{\left(\sum_{i=1}^{n_q} \boldsymbol{p}_i\right)^2}{2\sum_{i=1}^{n_q} m_i} + V,  \label{eq:hamiltonian} 
\end{align}
with
\begin{align}
V = \sum_{j > i = 1}^{n_q} \left( V_{ij}^{\rm ch} + V_{ij}^{\rm OGE} + V^{\rm conf}_{ij} \right), 
\end{align}
where $\boldsymbol{p}_i$ is the three-momentum of the $i$th constituent quark, and $n_q$ is the number of constituent quarks in the system under consideration. The term $\left( \sum_{i=1}^{n_q} \boldsymbol{p}_i \right)^2 / \left( 2\sum_{i=1}^{n_q} m_i \right)$ represents the kinetic energy of the c.m. motion. Note that the Hamiltonian in Eq.~\eqref{eq:hamiltonian} is unified for both single baryons and baryon-baryon systems in our framework.

\subsection{Determination of parameters}

The model parameters in the Hamiltonian of Eq.~\eqref{eq:hamiltonian} include: (1) the constituent masses of $u(d)$ quark, $m_u$, and of $s$ quark, $m_s$; (2) the masses of pseudoscalar mesons $m_\pi$, $m_K$, $m_\eta$, and $m_{\eta'}$, and scalar mesons $m_{\sigma'}$, $m_\kappa$, $m_\epsilon$, and $m_\sigma$; (3) the cutoff mass $\Lambda$ in the form factor of Eq.~\eqref{eq:FF}; (4) the quark and chiral field coupling constant $g_{\rm ch}$; (5) the OGE coupling constants $g_u$ and $g_s$; and (6) the parameters of the phenomenological confinement potential, namely, the confining strengths $a_{uu}$, $a_{us}$, and $a_{ss}$, and zero point energies $a^0_{uu}$, $a^0_{us}$, and $a^0_{ss}$. 

As mentioned in the Introduction, in previous quark model calculations, the oscillator frequencies in the Gaussian wave functions of constituent quarks [see Eqs.~\eqref{eq:wf_B_A} and \eqref{eq:wf-spatial-B}] were also treated as model parameters. They were typically assumed to be the same for all octet and decuplet baryons, with the common oscillator frequency $\omega$ predetermined by choosing a specific value for the size parameter $b_u$ for $u(d)$ quark, usually in the range of $0.45$-$0.55$ fm (note that $b^{-2}_u = m_u \omega$, and the size parameter for strange quark $b_s = \sqrt{m_u/m_s} \, b_u$).

In our opinion \cite{Huang:2018}, once the parameters in the Hamiltonian are fixed, the oscillator frequencies in the Gaussian wave functions of constituent quarks should no longer be treated as free parameters. Instead, they should be automatically determined by solving the variational Schr\"odinger equations. Only then can the single-baryon wave functions be considered optimal solutions (within the single-Gaussian ansatz) of the given Hamiltonian. This ensures that the results for $BB$ interactions are physically meaningful and the model parameters are properly determined, which in turn is expected to lead to more reliable predictions for multiquark systems.

\begin{table}[tbp]
\caption{\label{Table:para} Values of the adjustable model parameters. The $\sigma$-meson mass $m_\sigma$ and the zero-point energies $a^{0}_{uu}$, $a^{0}_{us}$, and $a^{0}_{ss}$ are in MeV. The confinement strengths $a_{uu}$, $a_{us}$, and $a_{ss}$ are in MeV/fm$^2$ for quadratic confinement and in MeV/fm for linear confinement, respectively. }
\begin{tabular*}{\columnwidth}{@{\extracolsep\fill}lrr}
\hline\hline
  & Model I & Model II   \\ 
  & ($r^2$ conf.) & ($r$ conf.)   \\
\hline
$m_\sigma$   & $625.0\pm0.2$ & $608.0\pm0.1$   \\
$g_u$     & $0.9786\pm0.0005$ & $1.0640\pm0.0005$   \\
$g_s$   & $1.0742\pm0.0008$ & $1.1630\pm0.0007$   \\
$a_{uu}$ & $56.2\pm0.1$ & $86.5\pm0.1$    \\
$a_{us}$  & $69.3\pm0.2$ & $90.4\pm0.2$    \\
$a_{ss}$  & $101.3\pm0.4$ & $100.4\pm0.4$   \\
$a^{0}_{uu}$   & $-36.6\pm0.1$ & $-51.4\pm0.1$   \\
$a^{0}_{us}$   & $-25.1\pm0.2$ & $-33.4\pm0.3$   \\
$a^{0}_{ss}$   & $-14.9\pm0.3$ & $-12.8\pm0.4$   \\
\hline\hline
\end{tabular*}
\end{table}

\begin{table*}[tbp]
\caption{ \label{Table:single-baryon} Calculated masses, size parameters, and oscillator frequencies for the octet- and decuplet-baryon ground states. The masses and oscillator frequencies are in MeV, and the size parameters are in fm.}
\begin{tabular*}{\textwidth}{@{\extracolsep\fill}lcccccccc}
\hline\hline
 & $N$ & $\Lambda$ & $\Sigma$ & $\Xi$ & $\Delta$ & $\Sigma^*$ & $\Xi^*$ & $\Omega$  \\ \hline
Experiment  & $939$ & $1116$ & $1193$ & $1318$ & $1232$ & $1385$ & $1533$ & $1672$ \\
Theory  & $939$ & $1116$ & $1193$ & $1318$ & $1232$ & $1385$ & $1533$ & $1672$ \\
$b_u$  \big($\omega_B$\big) ~ \big($r^2$ conf.\big) & $0.472$ ($558$) & $0.473$ ($556$) & $0.495$ ($508$) & $0.476$ ($549$) & $0.588$ ($360$) & $0.578$ ($372$) & $0.561$ ($395$) & $0.540$ ($427$) \\
\hspace{1.27cm} \big($r$ conf.\big) & $0.474$ ($554$) & $0.478$ ($544$) & $0.507$ ($484$) & $0.487$ ($525$) & $0.642$ ($302$) & $0.632$ ($311$) & $0.615$ ($329$) & $0.593$ ($354$) \\
\hline\hline
\end{tabular*}
\end{table*}

As described in Ref.~\cite{Huang:2018}, our model parameters are determined as follows. The constituent quark masses are set to $m_u=313$ MeV and $m_s=470$ MeV. The masses of pseudoscalar and scalar mesons are taken from experiment, except for the $\sigma$ meson mass $m_\sigma$, which is treated as an adjustable parameter to be fixed by reproducing the deuteron binding energy and $NN$ scattering phase shifts. Specifically, we use $m_{\sigma'} = m_{\kappa} = m_{\epsilon}=980$ MeV, $m_{\pi}=138$ MeV, $m_K=495$ MeV, $m_{\eta}=549$ MeV, and $m_{\eta'}=957$ MeV. The cutoff mass $\Lambda$ is set to $1100$ MeV. The quark and chiral-field coupling constant $g_{\rm ch}$ is determined by matching the $NN\pi$ coupling at the quark level to its empirical hadronic value:
\begin{equation}
\frac{g_{\rm ch}^2}{4\pi} = \frac{9}{25} \frac{m_u^2}{M_N^2} \frac{g_{NN\pi}^2}{4\pi},   \label{eq:gch}
\end{equation}
with ${g_{NN\pi}^2}/{4\pi}=13.67$. The remaining parameters---those in the OGE and confinement potentials---together with the oscillator frequencies (or equivalently, the size parameters) for various baryons, are determined simultaneously by requiring that the minima of the Hamiltonian matrix elements for three-quark systems match the experimental masses of the octet and decuplet baryon ground states. The resulting values for all adjustable parameters are summarized in Table~\ref{Table:para}, where models I and II correspond to the models with quadratic and linear confinement, respectively. The uncertainties of the fitted parameters are estimated from the error bars of the input data points—namely, the energies of baryon ground states, the binding energy of the deuteron, and the $NN$ scattering phase shifts. Since these quantities have only minor error bars, the uncertainties in the fitted parameters turn out to be rather small. 

The masses, size parameters, and oscillator frequencies for the baryon ground states, calculated using the parameters in Table~\ref{Table:para}, are listed in Table~\ref{Table:single-baryon}. Our calculated masses for all octet and decuplet baryons agree very well with the experimental data. The resulting size parameters and oscillator frequencies differ significantly among different baryons. In particular, the oscillator frequencies for decuplet baryons are considerably smaller than those for octet baryons. For example, under the quadratic confinement potential, the oscillator frequency for the $\Delta$ baryon is only about $2/3$ of that for the $N$ baryon; under the linear confinement potential, it is even less than $55\%$ of that for the $N$ baryon. This clearly demonstrates that using a common oscillator frequency for all octet and decuplet baryons, as done in previous quark model studies, is really inadequate, especially when studying systems composed of one octet baryon and one decuplet baryon.

In addition to the masses of octet- and decuplet-baryon ground states, the parameter sets in Table~\ref{Table:para} also reproduce the deuteron binding energy and the $NN$ scattering phase shifts up to a total angular momentum $J=6$ and mixing parameters for the relevant coupled partial waves quite satisfactorily, as shown in Ref.~\cite{Huang:2018}.

\section{Application to the $N\Delta$ system} \label{Sec:discuss}

\begin{figure}[htb]
\includegraphics[width=\columnwidth]{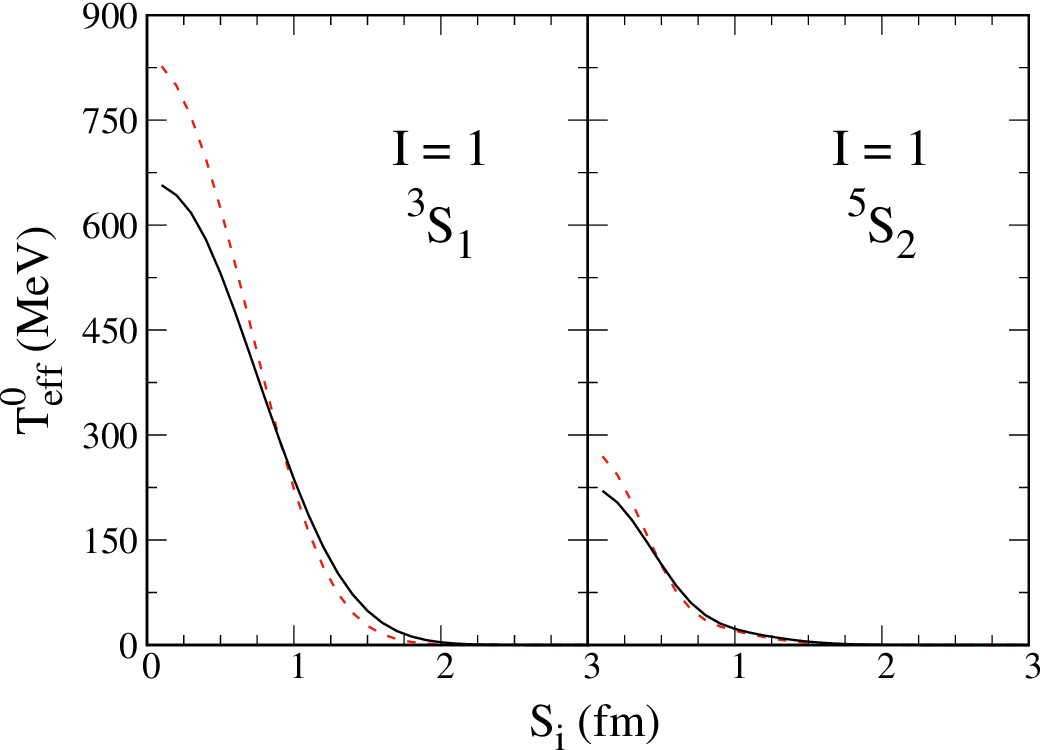}
\caption{Kinetic-energy contribution to the adiabatic interaction for the $S$-wave $N\Delta$ system (isospin $I=1$) as a function of the generator coordinate $S_i$ in model I. Solid lines: results from the present work. Dashed lines: results from traditional RGM calculations where the oscillator frequency for $\Delta$ is set equal to that for $N$.}
\label{fig:kine}
\end{figure}

\begin{figure}[htb]
\includegraphics[width=\columnwidth]{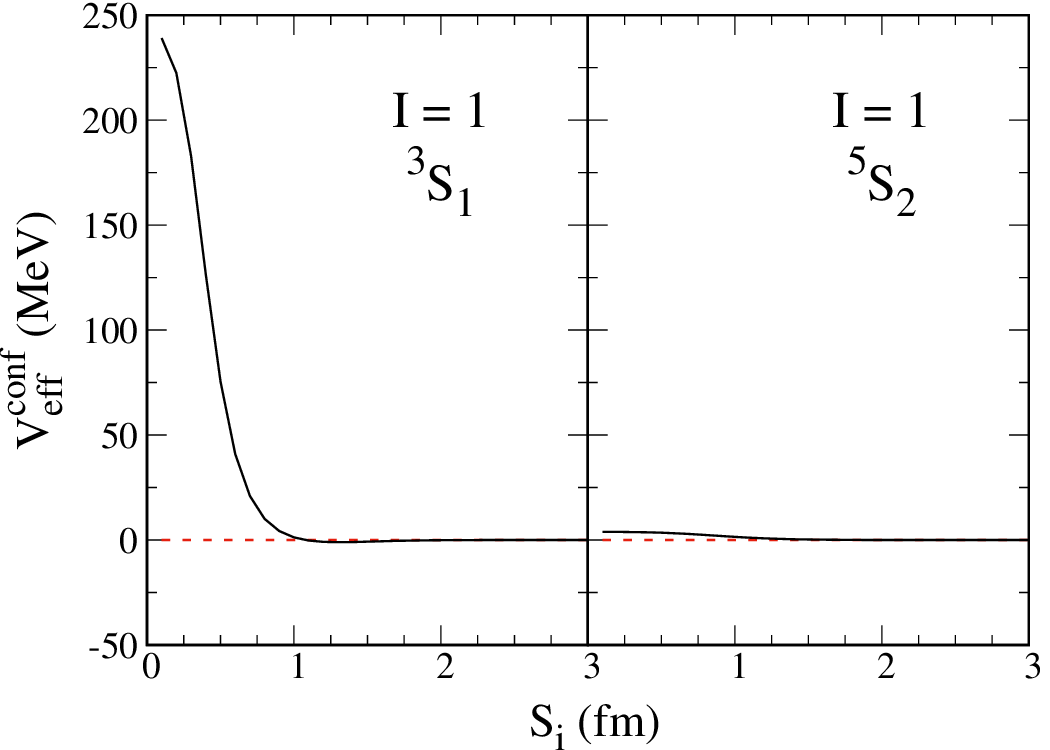}
\caption{Confinement-potential contribution to the adiabatic interaction for the $S$-wave $N\Delta$ system ($I=1$) in model I. Notations are the same as in Fig.~\ref{fig:kine}.}   \label{fig:conf}
\end{figure}

\begin{figure}[h!]
\includegraphics[width=\columnwidth]{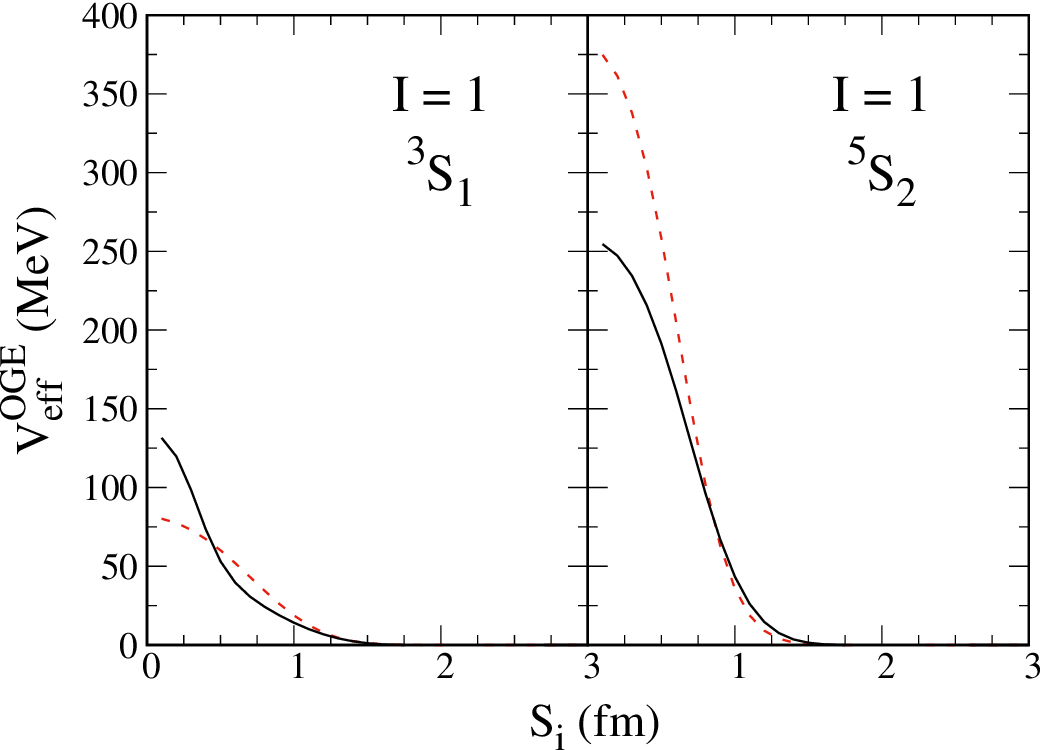}  \\[12pt] 
\includegraphics[width=\columnwidth]{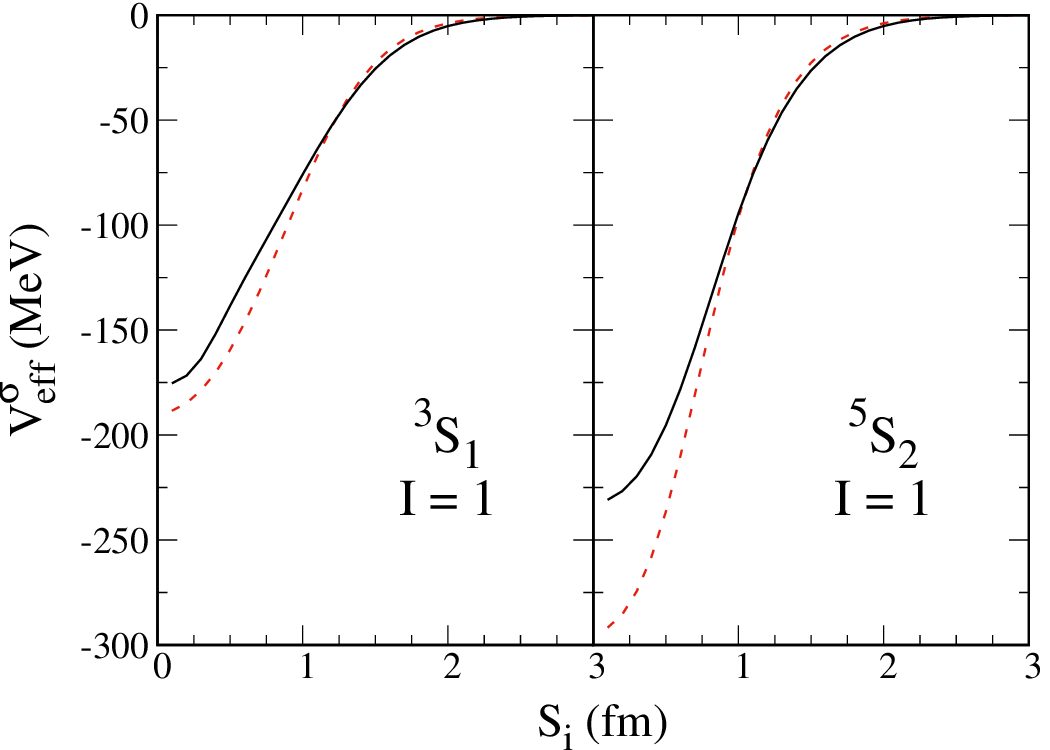}  \\[12pt] 
\includegraphics[width=\columnwidth]{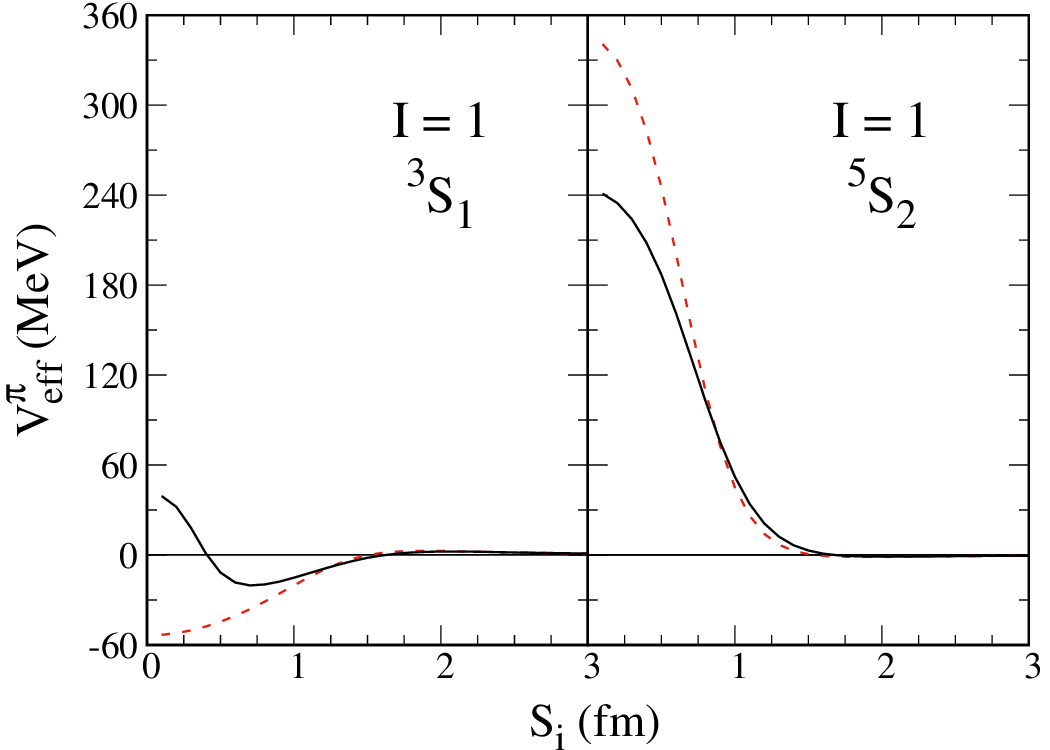}
\caption{Contributions from the OGE potential and from $\sigma$- and $\pi$-meson exchanges to the adiabatic interaction for the $S$-wave $N\Delta$ system ($I=1$) in model I. Notations are the same as in Fig.~\ref{fig:kine}.}   \label{fig:OGE-sig-pi}
\end{figure}

\begin{figure}[h!]
\includegraphics[width=\columnwidth]{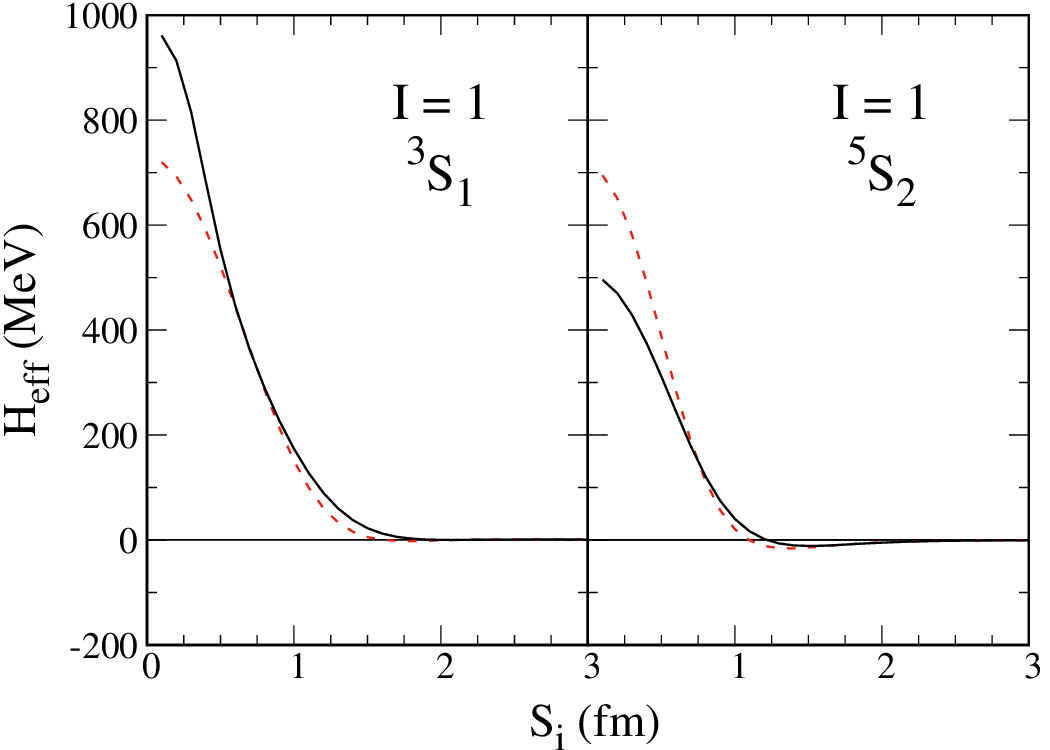}  \\[12pt] 
\includegraphics[width=\columnwidth]{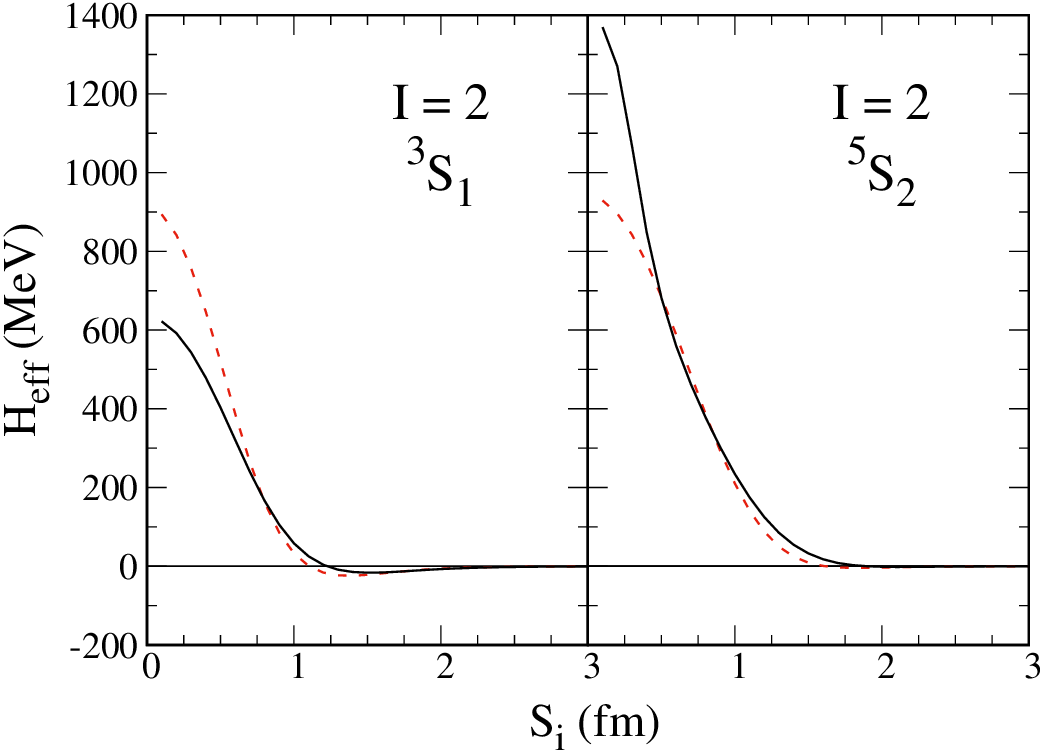}
\caption{Total Hamiltonian-induced adiabatic interactions for the $S$-wave $N\Delta$ systems with $I=1$ (upper panel) and $I=2$ (lower panel) in model I. Notations are the same as in Fig.~\ref{fig:kine}.}    \label{fig:H-eff}
\end{figure}

\begin{figure*}[htb]
\includegraphics[scale=0.5]{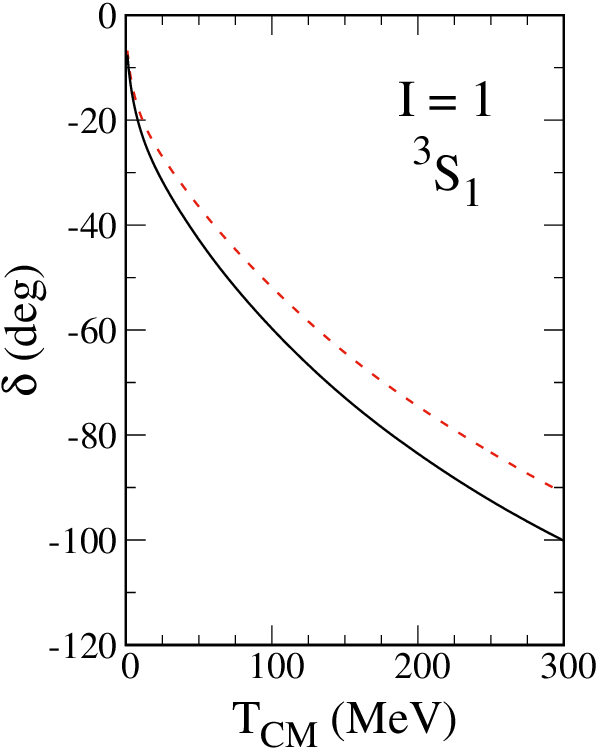} \hspace{1em}
\includegraphics[scale=0.5]{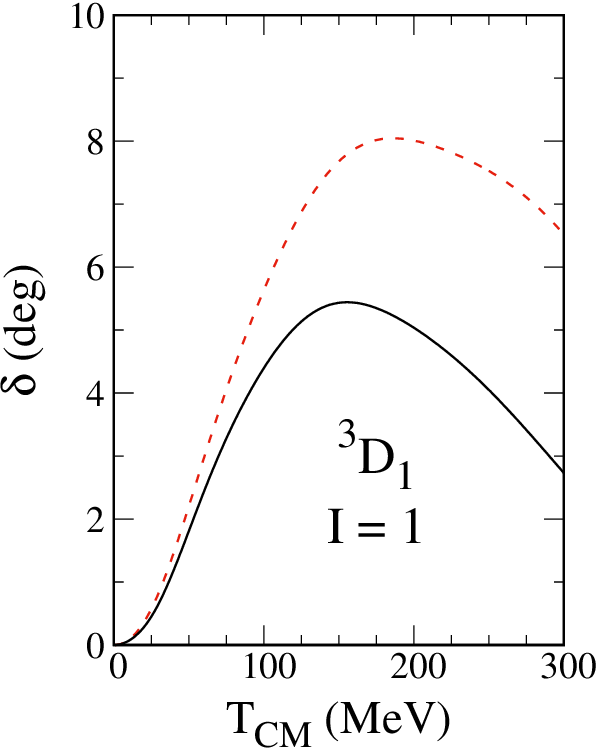}
\caption{$N\Delta$ ($I=1$) phase shifts for the $^3S_1$ and $^3D_1$ partial waves in model I. Notations are the same as in Fig.~\ref{fig:kine}.}   \label{fig:phase-3S1-3D1}
\end{figure*}

\begin{figure*}[htb]
\includegraphics[scale=0.5]{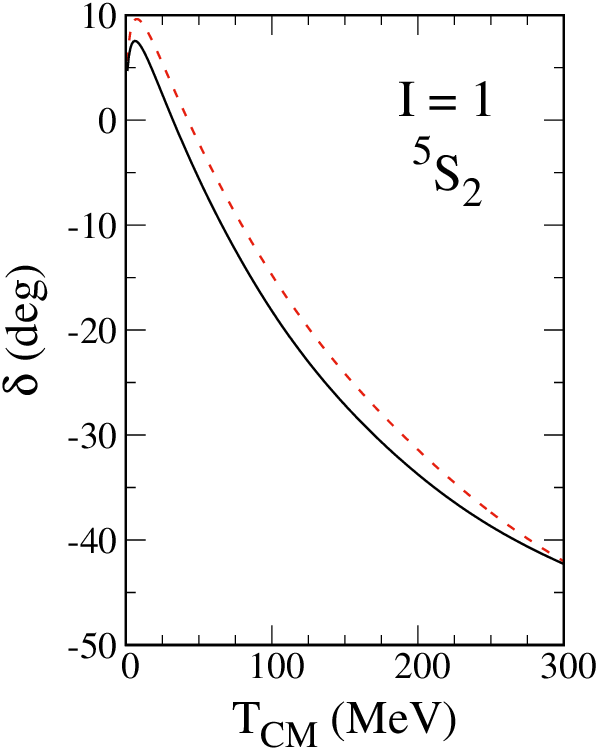} \hspace{1em}
\includegraphics[scale=0.5]{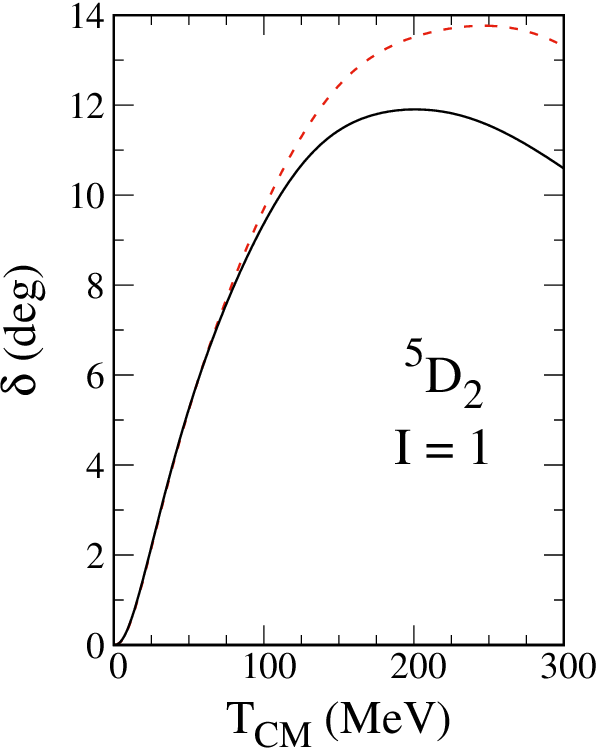} \hspace{1em}
\includegraphics[scale=0.5]{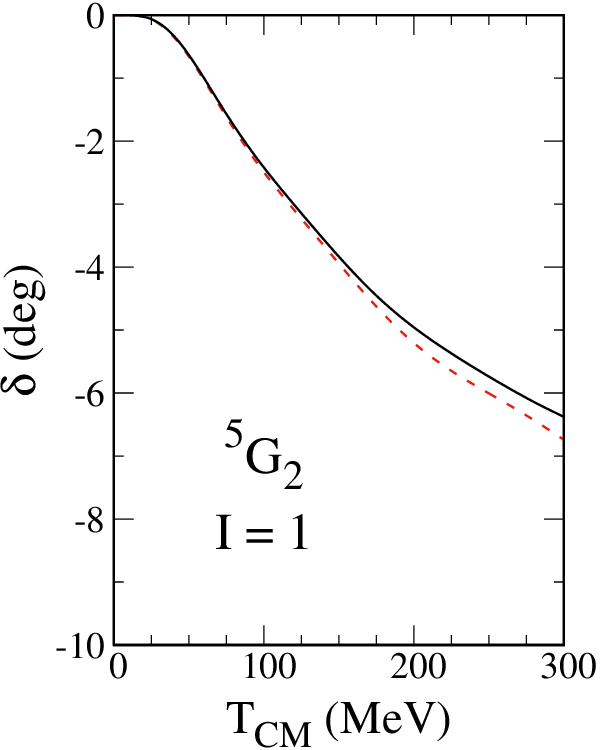}
\caption{$N\Delta$ ($I=1$) phase shifts for the $^5S_2$, $^5D_2$, and $^5G_2$ partial waves in model I. Notations are the same as in Fig.~\ref{fig:kine}.}   \label{fig:phase-5S2-5D2-5G2}
\end{figure*}

The new RGM formalism developed in Sec.~\ref{Sec:RGM} for two-baryon systems with unequal oscillator frequencies enables a consistent quark-level description of both single baryons and baryon-baryon interactions. Unlike previous quark-model investigations where all octet and decuplet baryons were assigned the same oscillator frequency, in our approach the oscillator frequency for each baryon is determined automatically by solving the variational Schr\"odinger equation. The new formalism correctly handles interactions between two baryons with distinct oscillator frequencies. In this section, we apply it to the $N\Delta$ system within the chiral SU(3) quark model, with the primary aim of highlighting the differences in the predicted $N\Delta$ interactions between the new formalism and traditional approaches. For this purpose, we present results only in the model with quadratic confinement potential (model I) in the main text; the corresponding results in the model with linear confinement potential (model II) are relegated to the Appendix, as they do not affect the qualitative conclusions regarding the new formalism.

In the RGM framework, the magnitude of the generator coordinate $\boldsymbol{S}_i$ corresponds to the distance between the centers of the two baryons [cf. Eq.~\eqref{eq:Si}]. At the outermost mesh point $\boldsymbol{S}_n$, where the two clusters are sufficiently separated, their mutual interaction is considered negligible. Therefore, the adiabatic interaction between the two baryons is conventionally defined as
\begin{align}
H^L_{\rm eff}(S_i) \equiv {\cal H}^{LL}_{ii}/{\cal N}^{LL}_{ii} - {\cal H}^{LL}_{nn}/{\cal N}^{LL}_{nn},
\end{align}
where the diagonal matrix elements ${\cal H}^{LL}_{ii}$ and ${\cal N}^{LL}_{ii}$ are given in Eq.~\eqref{eq:RGM-Lji}.

Figure~\ref{fig:kine} shows the kinetic-energy contribution to the adiabatic interaction for the $S$-wave $N\Delta$ system (isospin $I=1$) as a function of the generator coordinate $S_i$ in model I. There, solid and dashed lines represent results from the present work and from traditional RGM calculations (with $\omega_\Delta$ set equal to $\omega_N$), respectively. It is seen from this figure that the kinetic energy provides a repulsive contribution in both the $^3S_1$ and $^5S_2$ partial waves. However, at small $S_i$, the repulsion predicted by the present work is always weaker. In particular, the kinetic energy at $S_i\to 0$ is considerably smaller, and the slope of the curve---indicating the strength of the repulsive force---is also reduced. Actually, for the $S$-wave $N\Delta$ system with $\omega_\Delta = \omega_N$, the kinetic-energy-induced adiabatic interaction reads
\begin{align}
T^0_{\rm eff}(S_i) = \frac{\omega_N}{2} + \frac{3\omega_N x}{2} \left[-1 + \frac{i_1(3x) - 3 \braket{P_{36}} i_1(x)}{i_0(3x) - 9 \braket{P_{36}} i_0(x)} \right],    \label{eq:Teff}
\end{align} 
where $x\equiv m_u \omega_N S_i^2 / 4$, and $\braket{P_{36}}$ is the matrix element of the exchange operator $P_{36}$ [cf. Eq.~\eqref{eq:anti-symmtrizer}] in spin-flavor-color space. In the limit $S_i \to 0$, i.e. $x\to 0$,
\begin{align}
T^0_{\rm eff}(S_i) \to  \left\{
\begin{array}{lcl} 
\dfrac{3\omega_N}{2} - \dfrac{3\omega_N}{2} x,  &&  \left( \braket{P_{36}} = 1/9 \right)  \\[6pt]
\dfrac{\omega_N}{2} - \dfrac{3\omega_N}{2} x.  &&  \left( \braket{P_{36}} \neq 1/9 \right)
\end{array}
\right.
\end{align}
For the $I=1$ $N\Delta$ system, $\braket{P_{36}} = 1/9$ for the $^3S_1$ partial wave and $1/81$ for the $^5S_2$ partial wave. This naturally explains why $T^0_{\rm eff}(S_i)$ near $S_i \to 0$ is nearly three times larger in the $^3S_1$ channel than in the $^5S_2$ channel, while the slopes of $T^0_{\rm eff}(S_i)$ near $S_i \to 0$ for both partial waves are nearly identical. When $\omega_\Delta \neq \omega_N$, the expression of the kinetic-energy-induced adiabatic interaction becomes more complex,
\begin{align}
T^0_{\rm eff}(S_i) =& \, \frac{\omega_N}{2} k + \frac{3\omega_N x}{2} k^2 \left[-1 + \frac{i_1(3kx) - \cdots}{i_0(3kx) - \cdots} \right] \nonumber \\[5pt]
& ~ + {\rm terms ~ proportional ~ to ~} \left(\omega_\Delta - \omega_N \right)^2,   \label{eq:Teff-unequal}
\end{align} 
with $k=2\omega_\Delta/\left(\omega_\Delta + \omega_N\right)$. Setting $\omega_\Delta \to \omega_N$, one sees that $k\to 1$ and Eq.~\eqref{eq:Teff-unequal} recovers Eq.~\eqref{eq:Teff}. A  comparison of Eqs.~\eqref{eq:Teff-unequal} and \eqref{eq:Teff} shows that as $S_i \to 0$, both $T^0_{\rm eff}(S_i)$ and its slope in the case $\omega_\Delta \neq \omega_N$ are reduced by factors of $k$ $(\approx 0.78)$ and $k^2$ $(\approx 0.61)$, respectively, relative to the equal-frequency case. At larger $S_i$, Fig.~\ref{fig:kine} indicates that the kinetic energy induced effective $N\Delta$ interaction in the present work becomes more repulsive than in previous RGM studies. 

Figure~\ref{fig:conf} displays the confinement-potential induced adiabatic interaction for the $S$-wave $N\Delta$ system ($I=1$) as a function of the generator coordinate $S_i$ in model I. In previous quark-model investigations, the quadratic type confinement potential was believed to give no contribution to interactions between two color-singlet hadron clusters. However, this holds only under the assumption that the involved two baryons have the same oscillator frequencies. In the present work, with $\omega_\Delta \neq \omega_N$, the confinement potential induced adiabatic interaction between $N$ and $\Delta$ reads
\begin{widetext} 
\begin{align}
V^{{\rm conf},L}_{\rm eff}(S_i) =& \, - \frac{72}{m_u} \frac{\left(\omega_N - \omega_\Delta\right)^2}{\omega_N \omega_\Delta \left(\omega_N + \omega_\Delta\right)} \left. \frac{ \left[ \braket{P_{36}}_1 N^{\rm EX}_1 + 2 \braket{P_{36}}_2 N^{\rm EX}_2 \right]^L }{ \left[ N^{\rm D}_1 - 9 \braket{P_{36}}_1 N^{\rm EX}_1 - 9 \braket{P_{36}}_2 N^{\rm EX}_2 \right]^L } \right|_{S_j = S_i} \nonumber \\[3pt]
& \, + 12 \frac{\left(\omega_N - \omega_\Delta\right)^2}{\left(\omega_N + \omega_\Delta\right)^2}  \left. \frac{ \left[ \left( \boldsymbol{S}_i + \boldsymbol{S}_j \right)^2 \braket{P_{36}}_1 N^{\rm EX}_1 + \left( \boldsymbol{S}_i - \boldsymbol{S}_j \right)^2 \braket{P_{36}}_2 N^{\rm EX}_2 \right]^L }{ \left[ N^{\rm D}_1 - 9 \braket{P_{36}}_1 N^{\rm EX}_1 - 9 \braket{P_{36}}_2 N^{\rm EX}_2 \right]^L } \right|_{S_j = S_i},  \label{eq:Vconf_eff}
\end{align}
\end{widetext}
where $N^{\rm D}$ and $N^{\rm EX}$ denote direct and exchange terms of the normalization matrix elements, respectively; subscripts ``1" and ``2" refer to the $N\Delta \leftarrow N\Delta$ and $\Delta N \leftarrow N\Delta$ components, respectively; and the square brackets $[\cdots]^L$ indicate partial-wave projection into the $L$-th partial wave. One sees from Eq.~\eqref{eq:Vconf_eff} that the contributions of quadratic type confinement potential to $N\Delta$ interaction are proportional to $(\omega_N - \omega_\Delta)$, which explains naturally why in previous quark-model calculations the quadratic type confinement potential does not offer any contributions to the interaction between two color-singlet baryons. For the $N\Delta$ system with different oscillator frequencies, the confinement potential induced adiabatic interaction dominantly comes from the first term of Eq.~\eqref{eq:Vconf_eff}, and it contributes only at small $S_i$, since $N^{\rm D}$ is much larger than $N^{\rm EX}$ at large $S_i$. In the $S$-wave $N\Delta$ with $I=1$, the first term of Eq.~\eqref{eq:Vconf_eff} at $S_i \rightarrow 0$ for the $^5S_2$ partial wave is about $2\%$ of that for the $^3S_1$ partial wave, which is why a noticeable confinement contribution appears in Fig.~\ref{fig:conf} for $^3S_1$ but not for $^5S_2$. This nonzero contribution is a distinctive feature of the new RGM formalism developed in this work. In previous equal-frequency RGM calculations, the confinement potential gave zero contribution to interactions between two color-singlet clusters precisely because the two baryons were assigned the same oscillator frequency. Once unequal frequencies are properly incorporated, the antisymmetrization of the six-quark system no longer cancels the confinement contribution. As a result, the $BB$ interaction matrix elements acquire an explicit dependence on the confinement parameters---and hence on the difference between the two oscillator frequencies. Thus, within our new formalism, baryon-baryon interactions gain sensitivity to the phenomenological confinement potential, a sensitivity that was entirely absent in previous approaches. This opens up a new avenue for probing the confinement potential---a longstanding puzzle in strong-interaction physics.

Figure~\ref{fig:OGE-sig-pi} shows the OGE potential and the $\sigma$- and $\pi$-exchange potentials induced adiabatic interactions for the $S$-wave $N\Delta$ system ($I=1$) as a function of the generator coordinate $S_i$ in model I. Compared to the equal-frequency case, at small $S_i$ the present results show that: (1) the OGE provides more repulsion in the $^3S_1$ partial wave and less repulsion in the $^5S_2$ partial wave; (2) the $\sigma$ exchange provides less attraction in both the $^3S_1$ and the $^5S_2$ partial waves; and (3) the $\pi$ exchange provides repulsion instead of attraction in the $^3S_1$ partial wave and less repulsion in the $^5S_2$ partial wave. 

Figure~\ref{fig:H-eff} shows the total Hamiltonian-induced adiabatic interactions for the $S$-wave $N\Delta$ systems with $I=1$ (upper panel) and $I=2$ (lower panel) as a function of the generator coordinate $S_i$ in model I. One sees that for the $I=1$ $^3S_1$ and $I=2$ $^5S_2$ partial waves, the interaction at $S_i\rightarrow 0$ is more repulsive in the present work than in the equal-frequency case, primarily due to the strong repulsive confinement contribution in the present calculation with the new RGM formalism. For the $I=1$ $^5S_2$ and $I=2$ $^3S_1$ partial waves, the interaction at $S_i\rightarrow 0$ is less repulsive in the present work, owing to the reduced kinetic repulsion and weaker repulsions from OGE and meson exchanges (the confinement contribution here is negligible; cf. Fig.~\ref{fig:conf}).

The possible existence of $N\Delta$ bound states (dibaryons) has a long history. In 1964, based on SU(6) symmetry and a mass formula, Dyson and Xuong predicted $D_{12}$ and $D_{21}$ states near the $N\Delta$ threshold \cite{Dyson64}. Subsequent studies using meson-exchange models have yielded conflicting predictions, with some reporting bound states \cite{Arenhovel75,Arndt87,Hoshizaki92,Gal14,Adlarson18,Lu20} and others finding none \cite{Belyaev79}. Quark-model studies, including early RGM calculations hampered by the equal-frequency assumption, have also been inconclusive \cite{Mulders83,Huang18}. In the present work, we have searched for bound states in all partial waves up to $L=2$ (including $S$, $P$, and $D$ waves) and for both isospins $I=1,2$ by solving the variational RGM equation \eqref{eq:RGM-bound-2} within the chiral SU(3) quark model. No solution with $E < m_N + m_\Delta$ was found in any channel, indicating that the attractive forces, primarily from $\sigma$-exchange, are insufficient to form a bound state.

Figures \ref{fig:phase-3S1-3D1} and \ref{fig:phase-5S2-5D2-5G2} show the $N\Delta$ ($I=1$) scattering phase shifts for the $^3S_1$, $^3D_1$, $^5S_2$, $^5D_2$, and $^5G_2$ partial waves in model I. One sees that the phase shifts for all partial waves except $^5G_2$ calculated in the present work are significantly different from those in traditional equal-frequency RGM calculations. The phase shifts for $^3S_1$ partial wave are negative, indicating a repulsive interaction, consistent with what we observed in Fig.~\ref{fig:H-eff}. Similarly, the phase shifts for $^5S_2$ partial wave show attraction at long range (low energy) and repulsion at short range (higher energy), as exhibited in Fig.~\ref{fig:H-eff}. For higher partial waves, the differences of the phase shifts from the present work and traditional equal-frequency RGM calculations get pronounced at higher energies. In particular, for $^5G_2$, visible differences emerge only above 300 MeV, outside the plotted range of Figs.~\ref{fig:phase-3S1-3D1} and \ref{fig:phase-5S2-5D2-5G2}. 

In summary, the significant differences in both adiabatic potentials and scattering phase shifts clearly demonstrate that the approximation of equal oscillator frequencies for $N$ and $\Delta$ in earlier RGM calculations is fundamentally inadequate. For a consistent quark-level description of baryon-baryon interactions, it is essential to use distinct oscillator frequencies, particularly for systems involving both octet and decuplet baryons. This conclusion naturally generalizes to microscopic quark-model studies of meson-meson and baryon-meson interactions.

\section{summary and conclusions} \label{Sec:summary}

Understanding hadronic phenomena at the quark level is a fundamental goal in hadron physics, as hadrons are composed of quarks and gluons. Over the past few decades, the RGM has become one of the most widely used approaches for studying hadron-hadron interactions, particularly  $NN$ interactions, within constituent quark models. As first pointed out in our previous work \cite{Huang:2018}, earlier quark-model investigations of $BB$ interactions often assumed the same oscillator frequency for all baryons. This assumption leads to the fact that the wave functions for individual baryons are not solutions of the given Hamiltonian, thereby casting doubt on the reliability of the calculated $BB$ interactions. 

In Ref.~\cite{Huang:2018}, we carried out the first RGM study of the $NN$ interaction that is consistent with the ground-state energies of all octet and decuplet baryons in a chiral SU(3) quark model. That work demonstrated that different baryons have significantly different size parameters (or oscillator frequencies) in their Gaussian wave functions, due to their distinct quantum numbers and the resulting interactions. Notably, the oscillator frequencies for octet baryons are much bigger than those for decuplet baryons [see Table~\ref{Table:single-baryon}]. Consequently, any RGM study of $BB$ interactions at the quark level must properly handle different oscillator frequencies for different baryons. This is a challenging task since previous RGM formulations in constituent quark models were applicable only to systems where the two baryons shared the same oscillator frequency.

In the present work, we develop a new RGM formalism suitable for studying interactions between two baryons with unequal oscillator frequencies. We detail the construction of wave functions for such two-baryon systems with the CM motion properly removed, and describe how to solve the corresponding RGM equations for bound-state and scattering problems.

We apply this new RGM formalism to study the $N\Delta$ interaction within a chiral SU(3) quark model, comparing the results with those from earlier RGM calculations that assumed equal oscillator frequencies. The adiabatic $N\Delta$ interactions obtained with the new formalism differ significantly from previous results. In particular, the confinement potential---which was believed to give zero contribution to interactions between two color-singlet baryons $N$ and $\Delta$ in earlier RGM studies---is found to yield a considerable short-range effect. This makes the $BB$ interactions a valuable arena for probing the phenomenology of color confinement. The analysis yields no $N\Delta$ bound states, and $N\Delta$ scattering phase shifts are predicted. These phase shifts exhibit marked differences compared to those calculated using the previous RGM formalism. The significant differences in both adiabatic potentials and scattering phase shifts clearly demonstrate the inadequacy of the assumption of equal oscillator frequencies for various baryons in earlier RGM calculations of $BB$ interactions.

In future work, we will employ the approach presented herein to achieve a unified description of single-baryon properties, $NN$ interactions, and $YN$ interactions within chiral quark models. This should allow a more physically grounded determination of model parameters and pave the way for more reliable investigations of other $BB$ interactions as well as exotic hadronic states such as dibaryons and multiquark systems.

\begin{acknowledgments}
This work is partially supported by the National Natural Science Foundation of China under Grants No.~12575093 and No.~12175240, and the Fundamental Research Funds for the Central Universities.
\end{acknowledgments}

\appendix

\section{Results for $N\Delta$ system in the model with linear confinement potential}

\begin{figure}[tb]
\begin{center}
\includegraphics[scale=0.5]{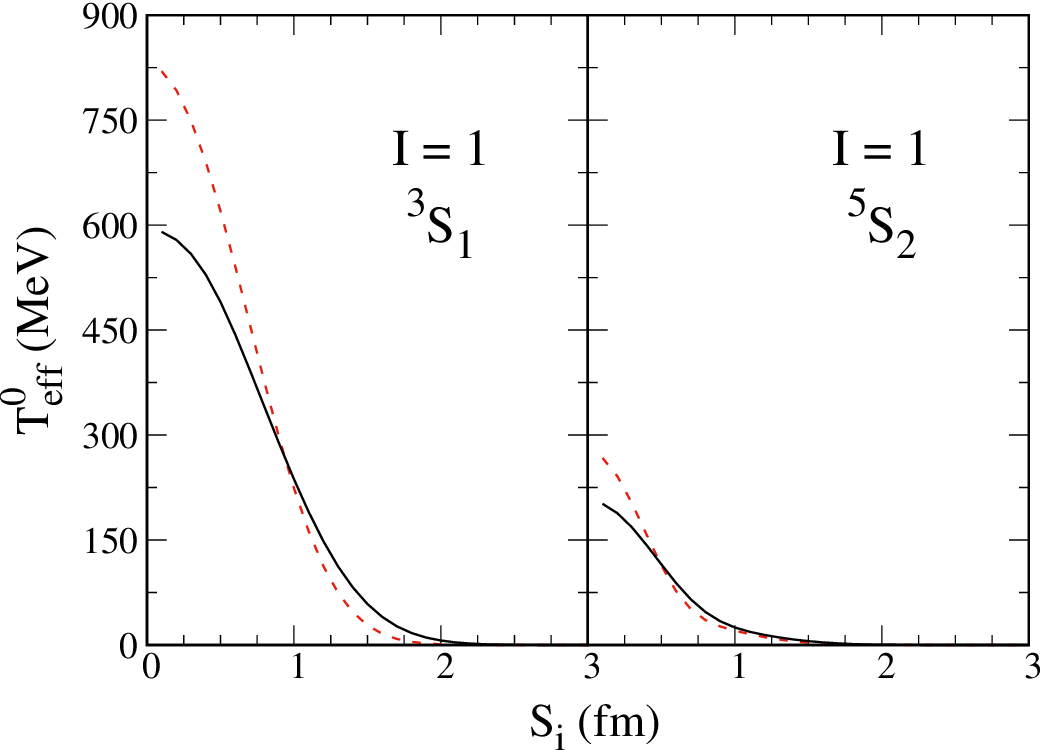}
\caption{Kinetic-energy contribution to the adiabatic interaction for the $S$-wave $N\Delta$ system (isospin $I=1$) as a function of the generator coordinate $S_i$ in model II. Notations are the same as in Fig.~\ref{fig:kine}.}
\label{fig:kine_lin}
\end{center}
\end{figure}

\begin{figure}[htb]
\begin{center}
\includegraphics[scale=0.5]{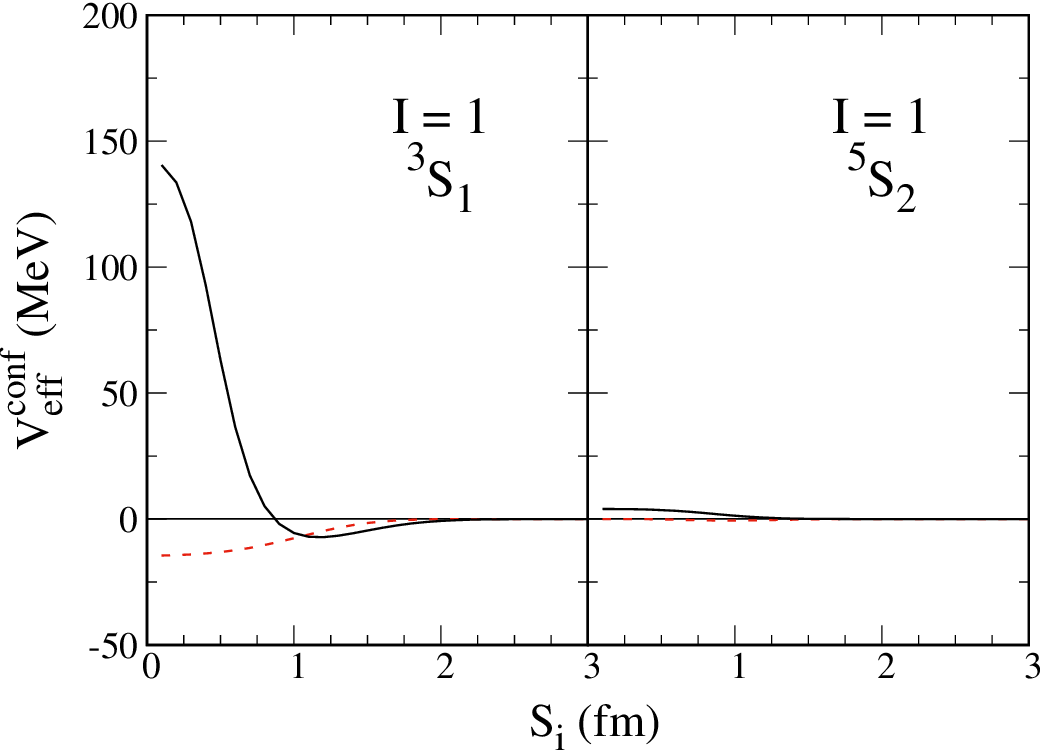}
\caption{Confinement-potential contribution to the adiabatic interaction for the $S$-wave $N\Delta$ system ($I=1$) in model II. Notations are the same as in Fig.~\ref{fig:kine}.}   \label{fig:conf_lin}
\end{center}
\end{figure}

\begin{figure}[h!]
\includegraphics[width=\columnwidth]{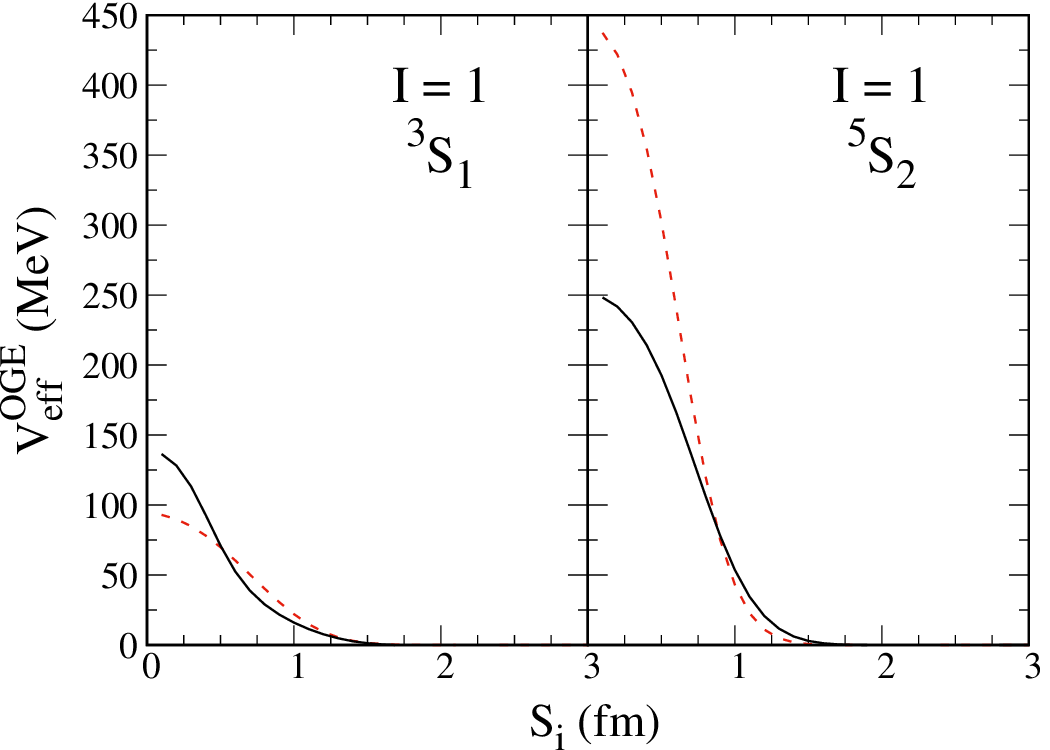}  \\[12pt] 
\includegraphics[width=\columnwidth]{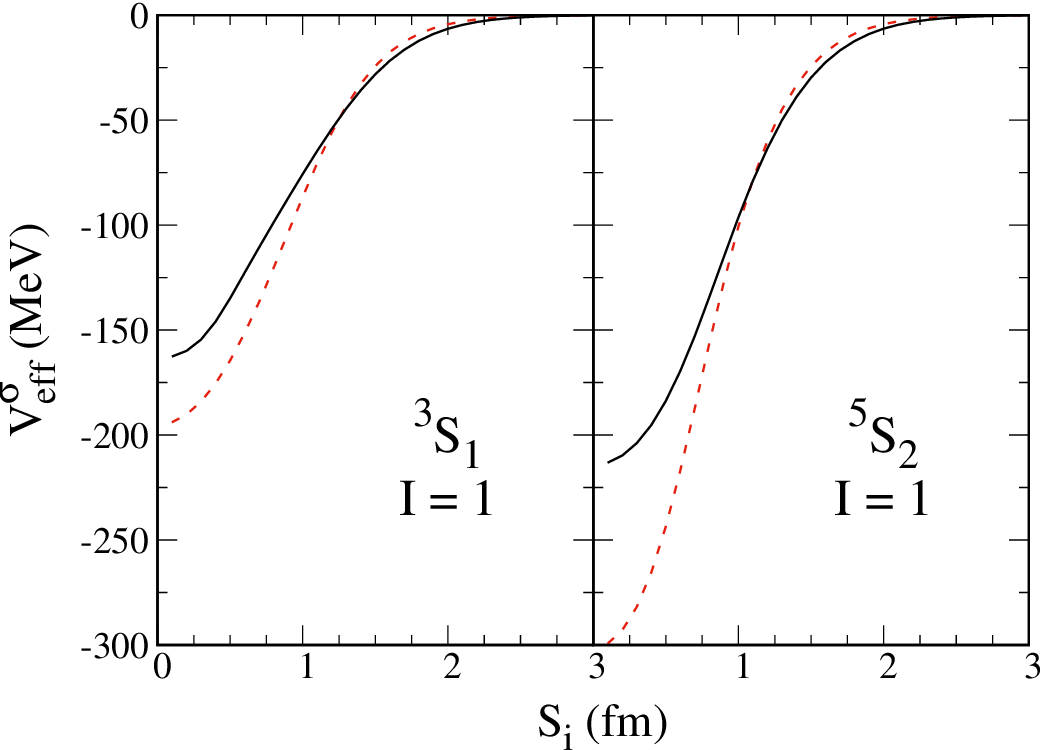}  \\[12pt] 
\includegraphics[width=\columnwidth]{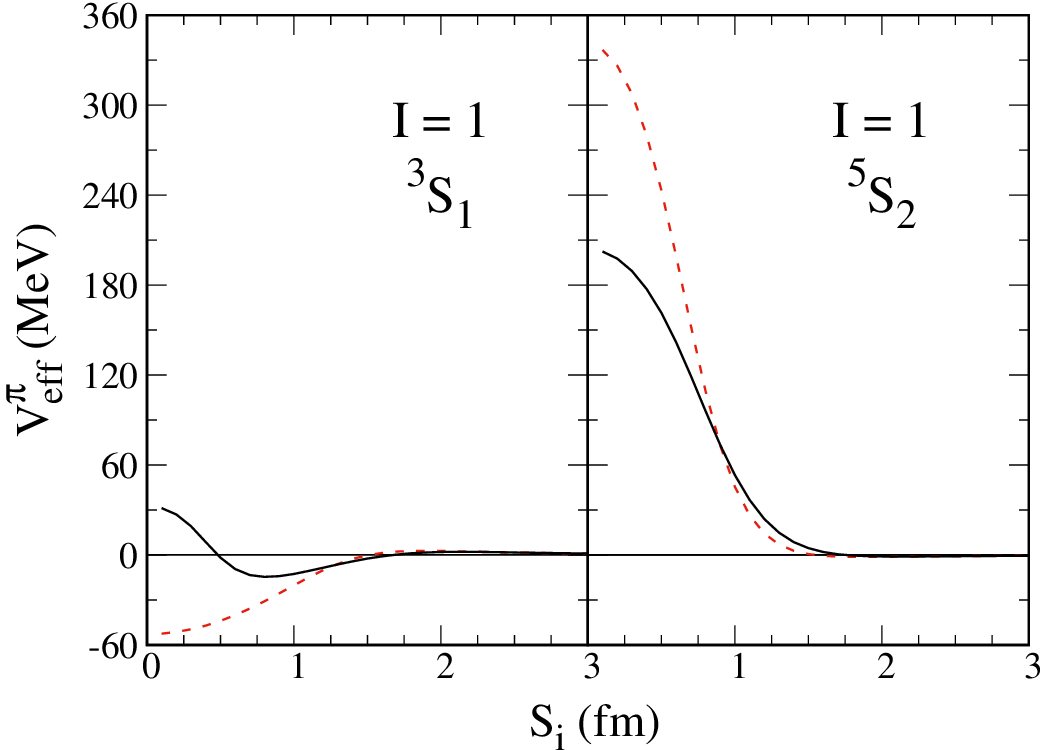}
\caption{Contributions from the OGE potential and from $\sigma$- and $\pi$-meson exchanges to the adiabatic interaction for the $S$-wave $N\Delta$ system ($I=1$) in model II. Notations are the same as in Fig.~\ref{fig:kine}.}   \label{fig:OGE-sig-pi_lin}
\end{figure}

\begin{figure}[htb]
\begin{center}
\includegraphics[scale=0.5]{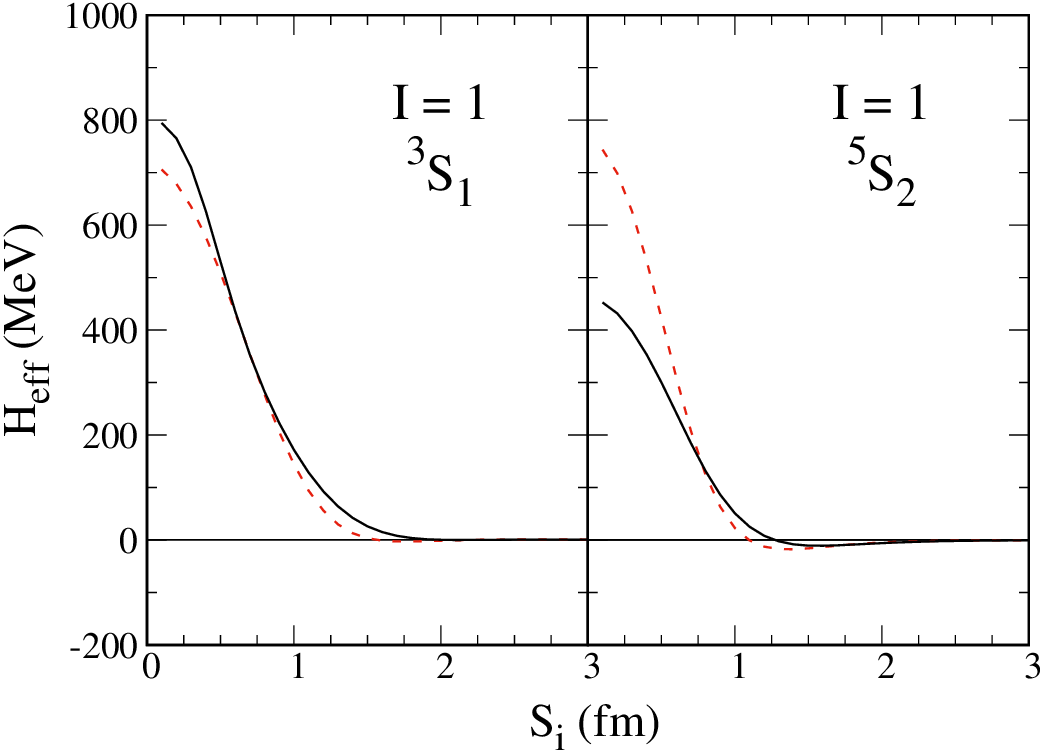} \\[12pt] 
\includegraphics[scale=0.5]{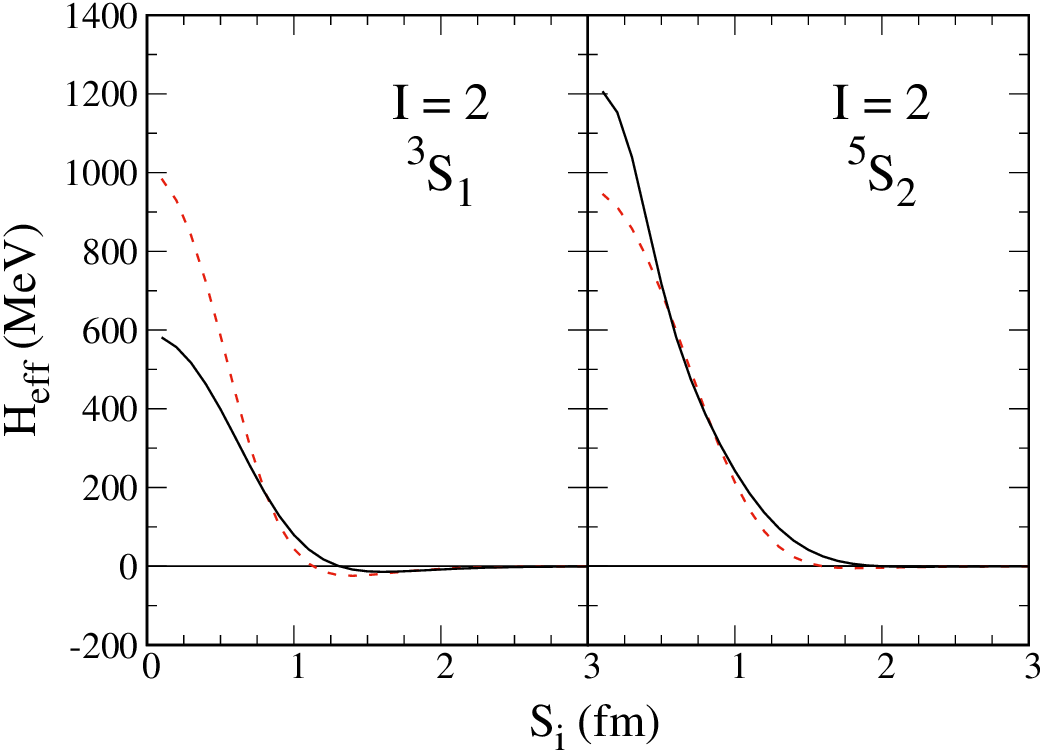}
\caption{Total Hamiltonian-induced adiabatic interactions for the $S$-wave $N\Delta$ systems with $I=1$ (upper panel) and $I=2$ (lower panel) in model II. Notations are the same as in Fig.~\ref{fig:kine}.}    \label{fig:H-eff_lin}
\end{center}
\end{figure}

\begin{figure*}[htb]
\begin{center}
\includegraphics[scale=0.5]{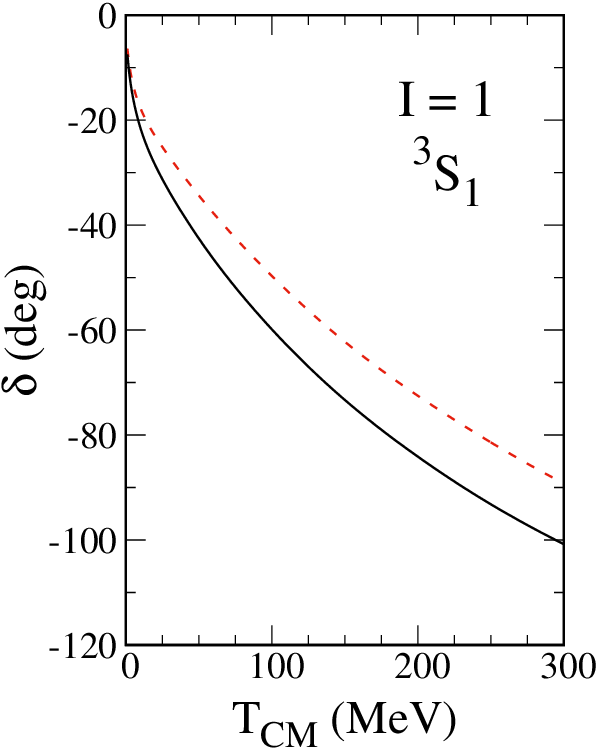} \hspace{1em}
\includegraphics[scale=0.5]{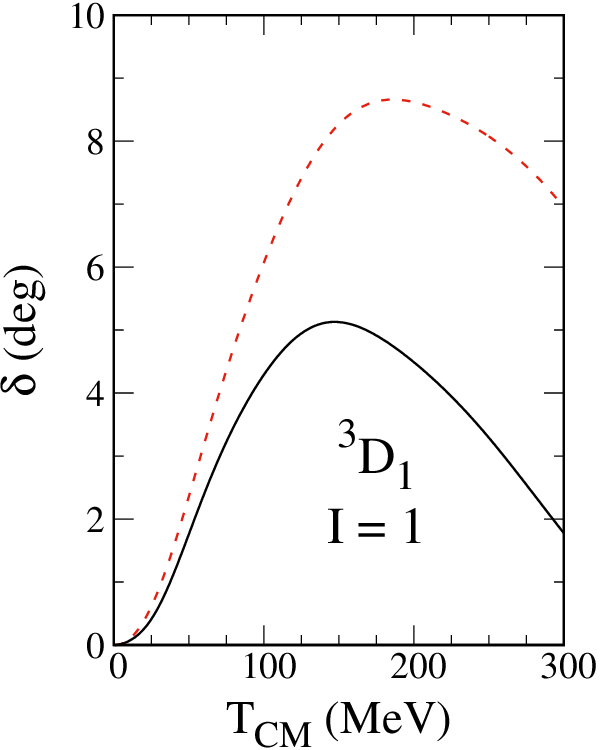}
\caption{$N\Delta$ ($I=1$) phase shifts for the $^3S_1$ and $^3D_1$ partial waves in model II. Notations are the same as in Fig.~\ref{fig:kine}.}   \label{fig:phase-3S1-3D1_lin}
\end{center}
\end{figure*}

\begin{figure*}[htb]
\begin{center}
\includegraphics[scale=0.5]{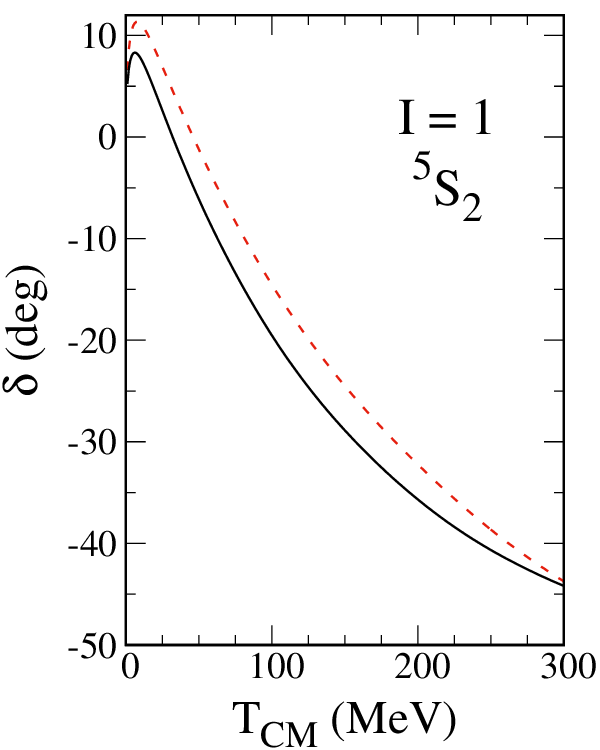} \hspace{1em}
\includegraphics[scale=0.5]{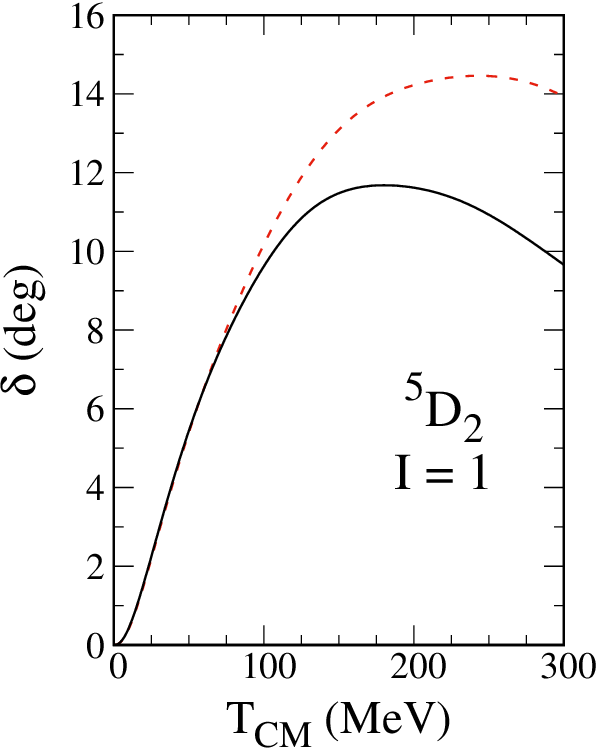} \hspace{1em}
\includegraphics[scale=0.5]{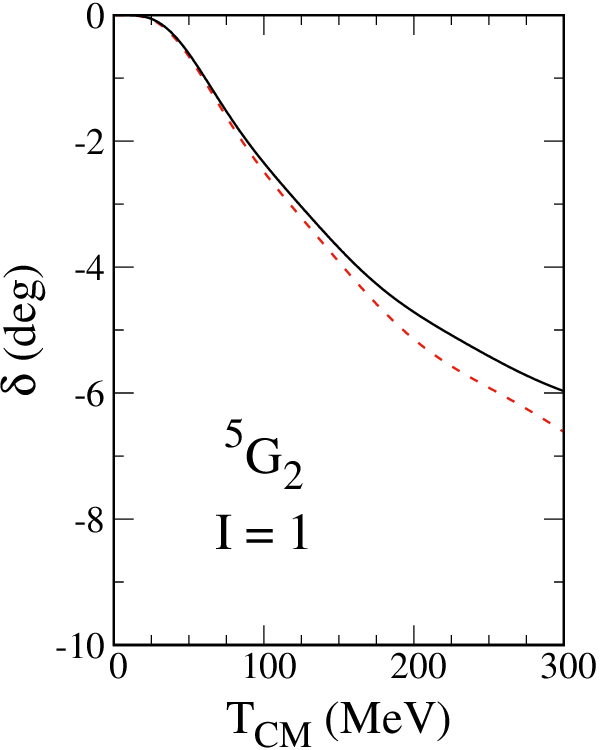}
\caption{$N\Delta$ ($I=1$) phase shifts for the $^5S_2$, $^5D_2$, and $^5G_2$ partial waves in model II. Notations are the same as in Fig.~\ref{fig:kine}.}   \label{fig:phase-5S2-5D2-5G2_lin}
\end{center}
\end{figure*}

In Sec.~\ref{Sec:discuss}, we presented and discussed the results for the $N\Delta$ system in the chiral SU(3) quark model with quadratic confinement potential (model I). In this appendix, we provide the corresponding results for model II, which employs a linear confinement potential, to further highlight the differences between the new RGM formalism (with different oscillator frequencies) and the traditional approach (with a common frequency). As explained in Sec.~\ref{Sec:discuss}, these results are relegated here because they do not affect the qualitative conclusions regarding the new formalism, but they are included to demonstrate that the main findings are robust against the choice of the confinement form. The general features of the results in model II are similar to those in model I, but with several important quantitative differences that we detail below.

Figure~\ref{fig:kine_lin} shows the kinetic-energy contribution to the adiabatic interaction for the $S$-wave $N\Delta$ system ($I=1$) as a function of the generator coordinate $S_i$ in model II. The notations are the same as in Fig.~\ref{fig:kine}: the solid and dashed lines represent results from the present work ($\omega_\Delta\ne \omega_N$) and from traditional RGM calculations ($\omega_\Delta = \omega_N$), respectively. Similar to the case of model I, the kinetic energy yields a repulsive effect. At short range ($S_i$ small), the new RGM formalism gives a significantly reduced kinetic energy and a milder slope compared with the equal-frequency case. At larger $S_i$, the kinetic-energy-induced effective $N\Delta$ interaction in the present work becomes more repulsive than in the traditional RGM calculations. Comparing the results shown in Figs.~\ref{fig:kine} and \ref{fig:kine_lin}, one sees that the kinetic-energy contributions in models I and II are the same in the traditional RGM calculations, since $\omega_\Delta$ is fixed to the same value regardless of the confinement type. In the new RGM calculations, however, the kinetic-energy contributions in model II are less repulsive than those in model I, because the oscillator frequency $\omega_\Delta$ in the linear confinement model is significantly smaller than that in the quadratic confinement model (cf. Table~\ref{Table:single-baryon}).
 
Figure~\ref{fig:conf_lin} displays the confinement-potential induced adiabatic interaction for the $S$-wave $N\Delta$ system ($I=1$) in model II. In previous quark model studies where $\omega_\Delta$ is set to $\omega_N$, the quadratic confinement potential does not contribute to the interactions between two color-singlet baryons at all, as proved by Eq.~\eqref{eq:Vconf_eff} and illustrated in Fig.~\ref{fig:conf}. However, as shown in Fig.~\ref{fig:conf_lin}, the linear confinement provides small but non-zero contributions to the $N\Delta$ interaction. In particular, it yields weak attraction in the $^3S_1$ partial wave. In the new RGM formalism, the linear confinement provides strong repulsion at short distances for the $^3S_1$ partial wave, with the strength of the repulsion weaker than that for quadratic confinement. It also yields weak attraction in the medium- and long-range regions, unlike the quadratic confinement potential which vanishes in the medium and long range. These features indicate that when different baryon sizes are properly taken into account, the confinement potential can contribute to $BB$ interactions in a way that is sensitive to its functional form---thus offering a potential discriminator between competing confinement models.

Figure~\ref{fig:OGE-sig-pi_lin} shows the adiabatic interactions induced by the OGE potential and by the $\sigma$- and $\pi$-exchange potentials for the $S$-wave $N\Delta$ system ($I=1$) in model II. The main qualitative features are similar to those in model I (Fig.~\ref{fig:OGE-sig-pi}): at small $S_i$, the OGE provides more repulsive in the $^3S_1$ channel and less in the $^5S_2$ channel compared with the equal-frequency case; the $\sigma$ exchange provides less attraction in both channels; and the $\pi$ exchange provides repulsion instead of attraction in the $^3S_1$ channel. Quantitatively, although the OGE coupling constant $g_u$ in model II is slightly larger than that in model I (cf. Table~\ref{Table:para}), the $\Delta$ size is also larger in model II; these two effects largely compensate each other, resulting in OGE contributions that are very similar in both models. A similar compensation occurs for $\sigma$ exchange: the smaller $\sigma$ mass in model II (608 MeV vs. 625 MeV in model I) offsets the larger $\Delta$ size, yielding comparable attractions. For $\pi$ exchange, the difference between the two models comes almost entirely from the larger $\Delta$ size in model II, which leads to slightly weaker repulsive interactions at short range.

Figure~\ref{fig:H-eff_lin} shows the total Hamiltonian-induced adiabatic interactions for the $S$-wave $N\Delta$ systems with $I=1$ (upper panel) and $I=2$ (lower panel) as a function of $S_i$ in model II. Similar to the case of model I, for the $I=1$ $^3S_1$ and $I=2$ $^5S_2$ partial waves, the short-range interaction is more repulsive in the present work than in previous quark model calculations, mainly due to the repulsion provided by the confinement potential in the present study with unequal oscillator frequencies for $N$ and $\Delta$. For the $I=1$ $^5S_2$ and $I=2$ $^3S_1$ partial waves, the short-range interaction is less repulsive in the present work, primarily owing to the reduced kinetic repulsion and weaker repulsions from OGE and meson exchanges. Comparing model II with model I, one sees that in the traditional calculations the $N\Delta$ interactions are nearly identical in the two models. In the present work, however, the former is considerably less repulsive than the latter at short range for the $I=1$ $^3S_1$ and $I=2$ $^5S_2$ channels, mainly because of the weaker kinetic and confinement repulsions at small $S_i$. For the $I=1$ $^5S_2$ and $I=2$ $^3S_1$ partial waves, the effective $N\Delta$ interactions in the two models are similar in the present work, since the kinetic energy contributions are close and the confinement contributions are negligible in both models for these channels. 

To check whether there is an $N\Delta$ bound state in model II, we solve the RGM equation for bound-state problems [cf. Eq.~\eqref{eq:RGM-bound-1}]. The results yield no bound state for any partial wave up to $L=2$ (including $S$, $P$, and $D$ waves) for either isospin. In fact, although for the $I=1$ $^3S_1$ and $I=2$ $^5S_2$ partial waves the kinetic energy and confinement potential provide a less repulsive interaction in model II than in model I, the attractions---dominantly provided by $\sigma$ exchange---are still insufficient to bind the $N\Delta$ system. 

To further test the robustness of the no-bound-state conclusion, we performed a systematic scan over the parameter space. Specifically, we varied $g_u$ from $0.8$ to $1.2$ in steps of $0.02$ and $m_\sigma$ from $530$ to $700$ MeV in steps of $10$ MeV, keeping all other parameters fixed at their central values. These ranges are intentionally chosen to be much wider than the uncertainties quoted in Table~\ref{Table:para}. For both confinement models and for all partial waves up to $L=2$, no bound state was found throughout this entire parameter range. This confirms that the absence of $N\Delta$ bound states is a robust feature of our formalism.

Figures \ref{fig:phase-3S1-3D1_lin} and \ref{fig:phase-5S2-5D2-5G2_lin} show the $N\Delta$ ($I=1$) scattering phase shifts for the $^3S_1$, $^3D_1$, $^5S_2$, $^5D_2$, and $^5G_2$ partial waves in model II. It is seen that for all partial waves, the phase shifts from the present calculations differ significantly from those from previous quark model studies, which clearly demonstrates the importance of a careful and consistent treatment of the oscillator frequencies for various baryons, especially for systems involving decuplet baryons. Comparing with the results from model I, one observes that the phase shifts in the two models are very close to each other. This is because the interactions in these two models differ mainly at short range (small $S_i$), whereas the scattering phase shifts are predominantly determined by the long-range part of the interaction, which is similar in both models.

\bibliographystyle{apsrev}
\bibliography{project}

\end{document}